\documentclass[12pt]{article}
\usepackage{a4wide,latexsym,graphicx,epsfig,psfrag}
\usepackage{amsmath}
\usepackage{cite}
\newcommand{\chpt}{ChPT}
\newcommand{\nftwo}{\ensuremath{n_f=2}}
\newcommand{\nfthree}{\ensuremath{n_f=3}}
\newcommand{\trf}[1]{\ensuremath{\left\langle #1 \right\rangle}}
\newcommand{\cL}{\mathcal{L}}
\newcommand{\Q}{{\cal Q}}
\newcommand{\dg}{\dagger}

\newcommand{\wh}{\widehat}
\newcommand{\no}{\nonumber}
\newcommand{\nn}{\nonumber \\}
\newcommand{\ol}{\overline}
\begin{document}
\parskip=3pt plus 1pt

\begin{titlepage}
\vskip 1cm
\begin{flushright}
LU TP 14-16\\
UWThPh-2014-10\\ 
\end{flushright}

\setcounter{footnote}{0}
\renewcommand{\thefootnote}{\fnsymbol{footnote}}




\vspace*{1.5cm}
\begin{center}
{\Large\bf Mesonic low-energy constants}
\\[25mm]

{\normalsize\bf Johan Bijnens$^{1}$ and Gerhard Ecker$^{2}$}\\[1.2cm]
\end{center}   
\begin{center}
${}^{1)}$ Department of Astronomy and Theoretical Physics, Lund University,
\\S\"olvegatan 14A, 
SE 223-62 Lund, Sweden \\[10pt]
${}^{2)}$ University of Vienna, Faculty of Physics,\\
Boltzmanngasse 5, A-1090 Wien, Austria 
\end{center}

\vspace*{2cm}

\begin{abstract}
\noindent
We review the status of the coupling constants of chiral Lagrangians
in the meson sector, the so-called low-energy constants (LECs). Special
emphasis is put on the chiral $SU(2)$ and $SU(3)$ Lagrangians for the
strong interactions of light mesons. The theoretical and experimental
input for determining the corresponding LECs is discussed. In the 
two-flavour sector, we review the knowledge of the $O(p^4)$ LECs from 
both continuum fits and lattice QCD analyses. For chiral $SU(3)$,
NNLO effects play a much bigger role. Our main new results are fits of
the LECs $L_i$ both at NLO and NNLO, making extensive use of the available
knowledge of NNLO LECs. We compare our results with
available lattice determinations. Resonance saturation
of LECs and the convergence of chiral $SU(3)$ to NNLO are 
discussed.  We also review the status of 
predictions for the LECs of chiral Lagrangians with dynamical photons
and leptons. 
\end{abstract}


\vfill

\setcounter{footnote}{0}
\renewcommand{\thefootnote}{\arabic{footnote}} 

\end{titlepage}

\tableofcontents

\section{INTRODUCTION}

Low-energy meson physics and the study of the strong interaction at
low energies underwent a phase transition in the theoretical
description with the introduction of Chiral Perturbation Theory 
(\chpt) in the early 1980s \cite{Weinberg:1978kz,Gasser:1983yg,Gasser:1984gg}.
It allowed the theory of the lightest hadrons, the
pions, kaons and eta, to be put on a solid theoretical footing.
The main idea is that rather than a perturbative expansion in a small
parameter like $\alpha$ or $\alpha_S$, there is a well-defined perturbation
theory as an expansion in orders of momenta and masses. \chpt\ was also the
prototype effective field theory, showing how to make sense of nonrenormalizable
theories in a well-defined fashion.

The predictions of \chpt\ are of twofold type. There are the loop contributions
at each order and the contributions that involve the parameters of the
higher-order Lagrangians. This review summarizes the present knowledge
of the values of these parameters. The standard name for these parameters
is low-energy constants (LECs). The main part of this review concerns
the LECs of two- and three-flavour mesonic \chpt\ in the isospin 
limit. Section \ref{sec:chpt} gives an overview of the Lagrangians
and serves to define our notation.

A first determination of the LECs was done
in the papers where they were introduced \cite{Gasser:1983yg,Gasser:1984gg}.
For those of the two-flavour or \nftwo\ case, the various LECs at
next-to-leading order (NLO) can be determined in a 
rather straightforward fashion.
The main analyses have been pushed to next-to-next-to-leading order (NNLO).
This is reviewed in Sec.~\ref{sec:nf2} where we discuss the theoretical
and experimental input to determine them.

The three-flavour or \nfthree\ coefficients were first determined
in \cite{Gasser:1984gg} by using large-$N_c$ arguments, the relation with the
\nftwo\ LECs, the pseudoscalar masses and $F_K/F_\pi$.
The next step was to determine them from $K_{l4}$ decays at NLO
\cite{Riggenbach:1990zp,Bijnens:1989mr}. The first attempt at adding 
higher-order effects in determining the \nfthree\ LECs was
Ref.~\cite{Bijnens:1994ie}. 
The first full calculations at NNLO in \nfthree\ mesonic \chpt\
appeared in the late 1990s and a first fit using these expressions for
the LECs was done in \cite{Amoros:2000mc,Amoros:2001cp}. At this level,
there was not sufficient information to really determine all LECs at
NLO directly from data, in particular $L_4$ and $L_6$ are very
difficult to obtain.
The underlying reason for this is discussed in Sec.~\ref{sec:largenc}.
Another difficulty is that quark masses and LECs cannot be disentangled
without using more information \cite{Kaplan:1986ru}. We fix this ambiguity
by using the quark mass ratio $m_s/\hat m$ as input.
More calculations became available and partial analyses were performed
but a new complete analysis was done in \cite{Bijnens:2011tb}.
The main improvement of the refitting done in this review over
 \cite{Bijnens:2011tb} is a more extensive use of knowledge of the
NNLO LECs as discussed in Sec.~\ref{sec:nf3fits}. A minor improvement
is the inclusion of some newer $K_{l4}$ data. The data and theoretical
input used in the \nfthree\ fits beyond that already used for the \nftwo\
results are described in Sec.~\ref{sec:input3}. Our fitting and the new
central values for the LECs are given in Sec.~\ref{sec:nf3fits}.
The evidence for resonance saturation of both NLO and NNLO LECs is
discussed in Sec.~\ref{sec:resonance}. 
The quality of the fits and the convergence of the chiral expansion
are discussed in Sec.~\ref{sec:convergence}.

The LECs that show up in extensions with dynamical photons and leptons
cannot be determined from phenomenology directly but need further treatment.
We collect the known results in Sec.~\ref{sec:photons} where we pay
close attention to the correct inclusion of short-distance contributions.
For those involving the weak nonleptonic interaction we
only give a short list of the main references in Sec.~\ref{sec:other}
and refer to the recent
review \cite{Cirigliano:2011ny} for more references and details. Likewise,
we remain very cursory with respect to the anomalous intrinsic parity
sector in Sec.~\ref{sec:other}.

Lattice QCD has started to make 
progress in the determination of LECs,
especially for those involving masses and decay constants.
We rely heavily on the flavour lattice averaging group (FLAG) reports
\cite{Colangelo:2010et,Aoki:2013ldr}. Specific results are quoted and
compared with our continuum results in Secs.~\ref{sec:nf2lattice}
and \ref{sec:nf3lattice}. Some comments can also be found in
Sec.~\ref{sec:largenc}.

A summary of the main results can be found in the conclusions.

\section{CHIRAL PERTURBATION THEORY}
\label{sec:chpt}

\chpt\ dates back to current algebra but its modern form was introduced
by the papers of Weinberg, Gasser and Leutwyler
\cite{Weinberg:1978kz,Gasser:1983yg,Gasser:1984gg}.
The underlying idea is to use the global chiral symmetry present in the
QCD Lagrangian for two (\nftwo) or three (\nfthree)
light quarks when the quark masses are put to zero. 
This symmetry is spontaneously broken in QCD. The
Nambu-Goldstone bosons resulting from this breaking are identified with the
pions (\nftwo) or the lightest pseudoscalar octet, $\pi$, $K$ and $\eta$
(\nfthree). The singlet axial symmetry is broken explicitly for QCD at the
quantum level due to the $U(1)_A$ anomaly and we thus disregard it. A
direct derivation of \chpt\ from the underlying assumptions 
is given by Leutwyler \cite{Leutwyler:1993iq}.

The perturbation in \chpt\ is not an expansion in a small
coupling constant but an expansion in momenta and quark masses.
Its consistency was shown in detail in \cite{Weinberg:1978kz} and is
often referred to as Weinberg or $p$ power counting.

A more extensive introduction to \chpt\ can be found in \cite{Scherer:2012xha}.
There are many reviews of \chpt. Those focusing on the meson
sector are two at the one-loop level \cite{Ecker:1994gg,Pich:1995bw}
and one at the two-loop level~\cite{Bijnens:2006zp}.

In terms of a quark field $\bar q =(\bar u~\bar d)$ (\nftwo) or
$\bar q =(\bar u~\bar d~\bar s)$ (\nfthree) the fermionic part of the
QCD Lagrangian can be written as
\begin{equation}
\label{QCD}
\mathcal{L}_{QCD} = \bar q i\gamma^\mu 
\left( \partial_\mu-ig_s G_\mu -i(v_\mu+\gamma_5 a_\mu)\right)q
 -\bar q s q+i\bar q p\gamma_5 q\,.
\end{equation}
The external fields or sources $v_\mu, a_\mu, s$ and $p$
are $n_f\times n_f$ matrices in flavour space.
They were introduced in \cite{Gasser:1983yg,Gasser:1984gg} to make chiral
symmetry explicit throughout the calculation and to facilitate the
connection between QCD and \chpt. For later use we define
$l_\mu=v_\mu-a_\mu$ and  $r_\mu = v_\mu+a_\mu$.

The degrees of freedom are the Goldstone bosons of
the spontaneous breakdown of the $SU(n_f)_L\times SU(n_f)_R$ chiral symmetry
of QCD with $n_f$ massless flavours to the vector subgroup $SU(n_f)_V$.
These are parametrized by a special $n_f\times n_f$ unitary matrix $U$.
The transformations under a chiral symmetry transformation 
$g_L\times g_R\in SU(n_f)_L\times SU(n_f)_R$ are
\begin{eqnarray}
U &\to& g_R U g_L^\dagger\,,{\qquad\qquad\qquad s+ i p\to g_R (s + i p)
g_L^\dagger}\,, \nonumber\\
l_\mu&\to& g_L l_\mu g_L^\dagger -i\partial_\mu g_L g_L^\dagger\,,\quad
r_\mu\to g_R r_\mu g_R^\dagger -i\partial_\mu g_R g_R^\dagger\,.
\end{eqnarray}
In addition we define $u$ with $u^2=U$ and $h(u,g_L,g_R)$ transforming
as
\begin{equation}
u\to g_R u h^\dagger = h u g_L^\dagger\,.
\end{equation}

The easiest way to construct Lagrangians is to use objects $X$ that transform
under chiral symmetry as $X\to h X h^\dagger$. For the present paper
these are $u_\mu,f_\pm^{\mu\nu}$, $\chi_\pm$ and $\chi_-^\mu$ defined by
\begin{eqnarray}
\label{eq:defs}
u_\mu &=& i\left[u^\dagger\left(\partial_\mu-ir_\mu\right)u
                -u\left(\partial_\mu-il_\mu\right)u^\dagger\right]\,,
\qquad \chi = 2B\left(s+ip\right)\,,
\nonumber\\
\chi_\pm &=& u^\dagger \chi u^\dagger\pm u\chi^\dagger u\,,
\qquad\qquad\qquad
\chi_-^\mu = u^\dagger D^\mu \chi u^\dagger - u D^\mu \chi^\dagger u\,,
\nonumber\\
f_\pm^{\mu\nu} &=& u F^{\mu\nu}_L u^\dagger \pm u^\dagger
F_R^{\mu\nu}u\,, \qquad\qquad ~~
D_\mu \chi =\partial_\mu \chi-i r_\mu \chi+i\chi l_\mu\,,
\nonumber\\
F_L^{\mu\nu}&=&\partial^\mu l^\nu-\partial^\nu l^\mu -i\left[l^\mu,l^\nu\right]\,,
\qquad
F_R^{\mu\nu}=\partial^\mu r^\nu-\partial^\nu r^\mu -i\left[r^\mu,r^\nu\right]\,.
\end{eqnarray}
The Lagrangian at lowest order, $p^2$, is known since long ago and is in the
present notation
\begin{equation}
\label{eq:LOlag}
\mathcal{L}_2 = \frac{F^2}{4}\trf{u_\mu u^\mu+\chi_+}
\end{equation}
where $\trf{\dots}$ denotes the $n_f$-dimensional flavour trace.
The notation we will use for \nftwo\ is $F$ and $B$, and for
\nfthree\ $F_0$ and $B_0$ for the constants in (\ref{eq:defs}) and
(\ref{eq:LOlag}).
The Lagrangians at next-to-leading order, $p^4$, were constructed
in \cite{Gasser:1983yg,Gasser:1984gg} for \nftwo\ and \nfthree.
The \nftwo\ Lagrangian is
\begin{eqnarray}
\label{eq:NLOlag2}
\mathcal{L}_4^{\nftwo}&=&
\frac{l_1}{4}\trf{u_\mu u^\mu}\trf{u_\nu u^\nu}
+\frac{l_2}{4}\trf{u_\mu u_\nu}\trf{u^\mu u^\nu}
+\frac{l_3}{16}\trf{\chi_+}^2 +\frac{i l_4}{4}\trf{u_\mu \chi_-^\mu}
\nonumber\\&&
+ \frac{l_5}{4}\trf{f_+^2-f_-^2}
+\frac{i l_6}{2}\trf{f_{+\mu\nu}u^\mu u^\nu}
- \frac{l_7}{16}\trf{\chi_-}^2
\nonumber\\&&
+ ~~{\rm 3 ~contact~terms} ~.
\end{eqnarray}

The \nfthree\ Lagrangian is
\begin{eqnarray}
\label{eq:NLOlag3}
\mathcal{L}_4^{\nfthree}&=&
 L_1\trf{u_\mu u^\mu}\trf{u_\nu u^\nu}
+L_2\trf{u_\mu u_\nu}\trf{u^\mu u^\nu}
+L_3\trf{u_\mu u^\mu u_\nu u^\nu}
\nonumber\\&&
+L_4\trf{u_\mu u^\mu}\trf{\chi_+}
+L_5\trf{u_\mu u^\mu\chi_+}
+L_6\trf{\chi_+}^2
+L_7\trf{\chi_-}^2
\nonumber\\&&
+\frac{L_8}{2}\trf{\chi_+^2+\chi_-^2}
-iL_9\trf{f_{+\mu\nu}u^\mu u^\nu}
+\frac{L_{10}}{4}\trf{f_+^2-f_-^2}
\nonumber\\&&
+ ~~{\rm 2 ~contact~terms} ~.
\end{eqnarray}

The Lagrangian for the general $n_f$-flavour case can be found in
\cite{Bijnens:1999sh}. Terms that vanish due to the equations of
motion have been dropped. 
This is discussed in detail in \cite{Bijnens:1999sh}.

The Lagrangians at $O(p^6)$ are of the form
\begin{equation}
\label{eq:NNLOlag}
\mathcal{L}_6^{\nftwo}=\sum_{i=1,56} c_i o_i\,,
\qquad
\mathcal{L}_6^{\nfthree}=\sum_{i=1,94} C_i O_i\,,
\end{equation}
The classification was done in \cite{Bijnens:1999sh}
after an earlier attempt \cite{Fearing:1994ga}. 
The form of the operators $o_i$ and $O_i$ can be found in 
\cite{Bijnens:1999sh}. In \cite{Haefeli:2007ty} an extra relation
for the \nftwo\ case was found reducing the number of terms there to 56.

Renormalization is done with a \chpt\ variant of $\overline{MS}$
introduced in \cite{Gasser:1983yg}. A detailed explanation valid
to two-loop order can be found in \cite{Bijnens:1997vq,Bijnens:1999hw}.

The relevant subtraction coefficients for all cases are known. 
These are then used to split the coupling constants in the Lagrangian into
an infinite and a renormalized part. This split is not unique, so below
is the definition of the renormalized constants that we use:
\begin{equation}
\label{eq:defLir}
\hat L_i = (c\mu)^{d-4}\left(\hat\Gamma_i\Lambda + \hat L_i^r(\mu)\right)\,.
\end{equation}
The divergent part is contained in $\Lambda=1/(16\pi^2(d-4))$
and in \chpt\ we use as a standard
$\ln c = -(1/2)\left(\ln 4\pi+\Gamma^\prime(1)+1\right)$.
For \nftwo , $\hat L_i=l_i$ and $\hat \Gamma_i=\gamma_i$
are derived and listed in
\cite{Gasser:1983yg}. For \nfthree, $\hat L_i=L_i$
and $\hat\Gamma_i = \Gamma_i$ are derived and listed in
\cite{Gasser:1984gg}. For \nfthree\ the convention is to directly list the
$L_i^r$ at a scale $\mu=0.77~$GeV. For \nftwo\ the convention is to quote
instead values for the $\mu$-independent $\bar l_i$ which are defined as
\begin{equation}
\label{eq:deflib}
\bar l_i = \frac{32\pi^2}{\gamma_i}l_i^r(\mu)-\ln\frac{M_\pi^2}{\mu^2}\,.
\end{equation}
The definition of the renormalized couplings at $O(p^6)$ is
\begin{equation}
\hat{C_i} =\frac{(c\mu)^{2(d-4)}}{\hat F^2}
\left(\hat C_i^r(\mu)-\hat \Gamma_i^{(2)}\Lambda^2-\left(\hat \Gamma_i^{(1)}
  +\hat \Gamma_i^{(L)}(\mu)\right)\Lambda\right)\,.
\end{equation}
The values for all quantities needed for the \nftwo\ and \nfthree\ cases
can be found in \cite{Bijnens:1999hw}. We will below quote the $c_i^r$ and
$C_i^r$ at a scale $\mu=0.77~$GeV and use the physical pion decay constant
$F_\pi=0.0922~$GeV to make the $\hat C_i^r$ dimensionless.

In addition to the operators listed explicitly in the Lagrangians 
(\ref{eq:LOlag},\ref{eq:NLOlag2},\ref{eq:NLOlag3},\ref{eq:NNLOlag})
there are also so-called contact terms. The corresponding
coefficients cannot be directly measured in physical quantities
involving mesons and are therefore not relevant for phenomenology. 
Nevertheless, they have in principle well-defined values from Green
functions of currents, but depend on the 
precise definitions of these currents.

\subsection{Other Lagrangians}
\label{sec:other}

In the treatment of radiative corrections for strong and semileptonic
processes at low energies, photons and leptons enter as dynamical
degrees of freedom. Consequently, additional effective Lagrangians are
needed. 

Leaving out the kinetic terms for photons and leptons, a single new
term arises to lowest order, $O(e^2 p^0)$ \cite{Ecker:1988te}:
\begin{equation} 
\label{eq:e2p0}
\mathcal{L}_{e^2p^0} = e^2 {F_0}^4 Z \langle \Q_L^{\rm em} \Q_R^{\rm
  em}\rangle ~.
\end{equation} 
The spurion fields
\begin{eqnarray}
\label{eq:QLQR}
\Q_L^{\rm em} =  u Q_L^{\rm em} u^\dg ~, \hspace*{.5cm} & \hspace*{.5cm}
\Q_R^{\rm em} = u^\dg Q_R^{\rm em} u  
\end{eqnarray} 
are expressed in terms of the quark charge matrix
\begin{eqnarray}  
\label{eq:Qem}
Q_L^{\rm em} = Q_R^{\rm em} &=& \left( \begin{array}{ccc} 2/3 & 0 & 0 \\ 0 & -1/3 & 0
    \\ 0 & 0 & -1/3 \end{array} \right) ~.
\end{eqnarray}  
The lowest-order electromagnetic LEC $Z$ can be determined either
directly from the pion mass difference ($Z \simeq 0.8$) or, in
principle more reliably, from a sum rule in the chiral limit 
($Z \simeq 0.9$) \cite{Moussallam:1998za}.  

Neglecting leptonic terms for the moment, the next-to-leading order
Lagrangian of $O(e^2 p^2)$ was constructed by Urech
\cite{Urech:1994hd}:  
\begin{eqnarray} 
\label{eq:Le2p2}
\cL_{e^2p^2} &=&
e^2 {F_0}^2 \left\{ \frac{1}{2} K_1 \; \langle (\Q^{\rm em}_L)^2 +
(\Q^{\rm em}_R)^2\rangle \; \langle u_\mu  
u^\mu\rangle   + K_2 \; \langle \Q^{\rm em}_L \Q^{\rm em}_R\rangle 
\; \langle u_\mu u^\mu \rangle \right. \no \\
&& \mbox{} - K_3 \; [\langle \Q^{\rm em}_L u_\mu\rangle 
\; \langle \Q^{\rm em}_L u^\mu
\rangle + \langle \Q^{\rm em}_R u_\mu\rangle 
\; \langle \Q^{\rm em}_R u^\mu\rangle ] \no \\
&& \mbox{} + K_4 \; \langle \Q^{\rm em}_L u_\mu\rangle 
\; \langle \Q^{\rm em}_R u^\mu \rangle
 + K_5 \; \langle[(\Q^{\rm em}_L)^2 + (\Q^{\rm em}_R)^2] 
u_\mu u^\mu\rangle \no \\
&& \mbox{} + K_6 \; \langle (\Q^{\rm em}_L \Q^{\rm em}_R + 
\Q^{\rm em}_R \Q^{\rm em}_L) u_\mu u^\mu\rangle 
 + \frac{1}{2} K_7 \; \langle (\Q^{\rm em}_L)^2 
+ (\Q^{\rm em}_R)^2\rangle \; \langle \chi_+\rangle
\no \\
&& \mbox{} + K_8\; \langle \Q^{\rm em}_L \Q^{\rm em}_R\rangle 
\; \langle \chi_+\rangle 
+ K_9 \; \langle [(\Q^{\rm em}_L)^2 + (\Q^{\rm em}_R)^2] 
\chi_+\rangle \no \\
&& \mbox{} + K_{10}\; \langle(\Q^{\rm em}_L \Q^{\rm em}_R 
+ \Q^{\rm em}_R \Q^{\rm em}_L) \chi_+\rangle 
 - K_{11} \; \langle(\Q^{\rm em}_L \Q^{\rm em}_R 
- \Q^{\rm em}_R \Q^{\rm em}_L) \chi_-\rangle \no \\
&& \mbox{}- iK_{12}\; \langle[(\wh \nabla_\mu \Q^{\rm em}_L) \Q^{\rm em}_L -
\Q^{\rm em}_L \wh \nabla_\mu \Q^{\rm em}_L 
- (\wh \nabla_\mu \Q^{\rm em}_R) \Q^{\rm em}_R + 
\Q^{\rm em}_R \wh \nabla_\mu \Q^{\rm em}_R] u^\mu\rangle \no \\
&&  \mbox{}+ K_{13} \; \langle (\wh \nabla_\mu \Q^{\rm em}_L) 
(\wh \nabla^\mu \Q^{\rm em}_R) \rangle  \no \\
&& \left. \mbox{} + K_{14} \; \langle (\wh \nabla_\mu \Q^{\rm em}_L) 
(\wh \nabla^\mu \Q^{\rm em}_L) +
(\wh \nabla_\mu \Q^{\rm em}_R) (\wh \nabla^\mu \Q^{\rm em}_R)\rangle  
\right\}, 
\end{eqnarray}  
where
\begin{eqnarray} 
 \label{eq:covder1}
\wh \nabla_\mu \Q^{\rm em}_L &=& 
u (D_\mu Q_L^{\rm em}) u^\dg , \no \\
\wh \nabla_\mu \Q^{\rm em}_R &=& 
u^\dg (D_\mu Q^{\rm em}_R) u ,
\end{eqnarray} 
with
\begin{eqnarray}  
\label{eq:covder2}
D_\mu Q^{\rm em}_L &=& \partial_\mu Q^{\rm em}_L 
- i[l_\mu,Q^{\rm em}_L], \no \\
D_\mu Q^{\rm em}_R &=& \partial_\mu Q^{\rm em}_R - i[r_\mu,Q^{\rm em}_R]~.
\end{eqnarray} 
In the presence of dynamical photons and leptons, the external fields
$l_\mu,r_\mu$ are modified as
\begin{eqnarray} 
\label{eq:lrmu}
l_\mu & \longrightarrow & v_\mu - a_\mu - e Q_L^{\rm em} A_\mu + \sum_{\ell=e,\mu}
(\bar \ell \gamma_\mu \nu_{\ell L} Q_L^{\rm w} + \ol{\nu_{\ell L}} 
\gamma_\mu \ell Q_L^{{\rm w}\dg}), \no \\
r_\mu & \longrightarrow & v_\mu + a_\mu - e Q_R^{\rm em} A_\mu
\end{eqnarray} 
where $A_\mu$ is the photon field and the weak charge matrix is
defined as
\begin{eqnarray}   
\label{eq:Qw}
Q_L^{\rm w} = - 2 \sqrt{2}\; G_F \left( \begin{array}{ccc}
0 & V_{ud} & V_{us} \\ 0 & 0 & 0 \\ 0 & 0 & 0 \end{array} \right) ~, &\qquad
\Q_L^{\rm w} = u Q_L^{\rm w}  u^\dg~.
\end{eqnarray} 
$G_F$ is the Fermi coupling constant and $V_{ud}$, $V_{us}$ are
Kobayashi-Maskawa matrix elements. 

For radiative corrections in semileptonic processes, one needs in
addition the leptonic Lagrangian \cite{Knecht:1999ag} 
\begin{eqnarray}  
\label{eq:Llept}
\cL_{\rm lept} &=& 
e^2 \sum_{\ell} \left \{ {F_0}^2 \left[  
X_1 \ol{\ell} \gamma_\mu \nu_{\ell L} 
\langle u^\mu  \{ \Q_R^{\rm em}, \Q_L^{\rm w} \} \rangle 
\right. \right. \nn 
&&  \left. \left.
+ X_2 \ol{\ell} \gamma_\mu \nu_{\ell L} 
\langle u^\mu  [\Q_R^{\rm em}, \Q_L^{\rm w}] \rangle
+ X_3 m_\ell \ol{\ell} \nu_{\ell L} \langle \Q_L^{\rm w} \Q_R^{\rm em} \rangle
\right. \right. \nn
&&  \left. \left.
+ i X_4 \ol{\ell} \gamma_\mu \nu_{\ell L} 
\langle \Q_L^{\rm w} \wh \nabla^\mu  \Q_L^{\rm em} \rangle
+ i X_5 \ol{\ell} \gamma_\mu \nu_{\ell L} 
\langle \Q_L^{\rm w} \wh \nabla^\mu  \Q_R^{\rm em} \rangle 
+ h.c. \right]  \right. \nn
&&  \left.
+ X_6 \bar \ell (i \! \not\!\partial + e \! \not\!\!A )\ell
+ X_7 m_\ell \ol \ell  \ell \right \}. 
\end{eqnarray} 
Estimates of the electromagnetic LECs $K_i$ and $X_i$ will be reviewed
in Sec.~\ref{sec:photons}.

In this review we restrict ourselves to mesonic Lagrangians
for strong and semileptonic processes including radiative
corrections. \chpt\ has been applied to many more 
cases even in the meson sector, which will not be treated in any
detail here. Neglecting lattice actions altogether, two more classes
of chiral Lagrangians have been considered for the treatment of
odd-intrinsic-parity (anomalous) processes and of nonleptonic
decays. We include some relevant references here for the Lagrangians
and for estimates of the corresponding LECs. 

Anomalous processes start at $O(p^4)$. The odd-intrinsic-parity
Lagrangian of $O(p^4)$ is given by the 
Wess-Zumino-Witten Lagrangian \cite{Wess:1971yu,Witten:1983tw}
that has no free parameters. The anomalous
Lagrangian of $O(p^6)$ has 23 LECs
\cite{Ebertshauser:2001nj,Bijnens:2001bb}.  
Only partial results are available for the numerical values of those
constants. The most promising approach is based on a short-distance
analysis with or without chiral resonance Lagrangians
\cite{Moussallam:1997xx,Knecht:2001xc,RuizFemenia:2003hm}. Electromagnetic
corrections for anomalous processes ($n_f=2$) have also been
investigated \cite{Ananthanarayan:2002kj}. 

The chiral Lagrangian for nonleptonic interactions of lowest order,
$O(G_F p^2)$, contains two LECs $g_8,g_{27}$ \cite{Cronin:1967jq}. The most
recent evaluation of these LECs, including isospin breaking
corrections, can be found in the review \cite{Cirigliano:2011ny}. 

The LECs of $O(G_F p^4)$ (22 couplings $N_i$ in the octet and 28
couplings $D_i$ in the 27-plet Lagrangians)
\cite{Kambor:1989tz,Ecker:1992de} are less known than
their strong counterparts at $O(p^4)$.  
The most recent phenomenological analysis of those combinations that 
occur in the dominant $K\to 2 \pi, 3 \pi$ decays can be found in
Ref.~\cite{Bijnens:2004ai}. Some of the LECs appearing in rare $K$
decays have also been analysed \cite{Cirigliano:2011ny}.

Resonance saturation of weak LECs
\cite{Ecker:1992de,D'Ambrosio:1997tb} suffers from the drawback that
the weak resonance couplings are unknown and that 
short-distance constraints are missing. Nevertheless, resonance
saturation provides at least a possible parametrization of the
LECs. The most systematic approach is based on
factorization (valid to leading order in $1/N_c$)
\cite{Pallante:2001he,Cirigliano:2003gt}, but 
higher-order corrections in $1/N_c$ may well be sizeable.

A different approach is to use the $1/N_c$ arguments in conjunction with the
underlying short-distance physics. This approach was pioneered by
\cite{Bardeen:1986uz} and further pursued in \cite{Bijnens:1998ee}.
A more recent discussion is in \cite{Buras:2014maa}.
One main problem here is to make sure that short- and long-distance
matching is performed in a clean fashion \cite{Bijnens:1999zn,Peris:1998nj}.

Finally, there is a chiral Lagrangian for electromagnetic corrections
to nonleptonic weak processes. The single LEC of lowest order,
$O(G_F e^2 p^0$), related to the electromagnetic penguin contribution
\cite{Bijnens:1983ye,Grinstein:1985ut}, is reasonably well known
\cite{Cirigliano:1999hj,Bijnens:2001ps,Cirigliano:2003gt}. 
The additional 14 LECs of
$O(G_F e^2 p^2)$ \cite{Ecker:2000zr} have again been estimated to leading
order in $1/N_c$ (factorization). In this way, the LECs can be
expressed in terms of Wilson coefficients, the strong LECs $L_5,L_8$
and the electromagnetic LECs $K_i$
\cite{Pallante:2001he,Cirigliano:2003gt}. 

\subsection{Contributions at each order and terminology}
\label{sec:terminology}

\chpt\ has been an active field since the early 1980s. As a
consequence, the same quantities are often denoted by different
symbols and terminology. The actual constants are referred to as
low-energy constants, parameters or coupling constants in the
Lagrangians and sometimes even referred to as counterterms.
For the purpose of this review, these are all equivalent.

Another source of confusion is the nomenclature used for the orders.
The use in this review is that the contribution to 
tree-level diagrams from $O(p^2)$ Lagrangians only is called lowest order
or order $p^2$ or tree-level. The next order, which consists of
tree-level 
diagrams with one vertex of the $O(p^4)$ Lagrangian and the remaining
vertices from the $O(p^2)$ Lagrangian and of one-loop diagrams
with only $O(p^2)$ vertices, is called order $p^4$ or next-to-leading order
or one-loop order. In the same vein, the third order is called
next-to-next-to-leading order or $p^6$ or two-loop order.

When adding other Lagrangians, one needs in addition to specify to
which order one has included electromagnetic or weak coupling
constants and, if applicable, $m_d-m_u$.

\section{TWO FLAVOURS}
\label{sec:nf2}

\subsection{Continuum input}
\label{sec:input2}

If one looks at the \nftwo\ Lagrangians we have two parameters at LO,
$F$ and $B$, $7+3$ at NLO and $52+4$ at NNLO. The $i+j$ notation refers to the
number of LECs and the number of contact terms.
For most processes of interest, the tree-level contributions at NNLO are small,
since the relevant scale in most of these cases is $M_\pi$.

This has the advantage that the determination of the NLO LECs does not depend
on how well we know the values of the $c_i^r$ but it makes
  comparison with models for the $c_i$ more uncertain.

All observables needed are known to NNLO.
The NLO results are all present in \cite{Gasser:1983yg}. The NNLO results for
$M_\pi^2$ and $F_\pi$ were done in \cite{Burgi:1996qi,Bijnens:1997vq}.
$\pi\pi$ scattering was done to NNLO in \cite{Bijnens:1995yn,Bijnens:1997vq}.
Finally, the scalar and vector form factors of the pion were obtained to two-loop
order in \cite{Bijnens:1998fm} while the pion radiative decay can be found in
\cite{Bijnens:1996wm}.

The pion mass is experimentally very well known \cite{Beringer:1900zz}. 
The larger question here is which pion mass to use, charged or neutral.
When comparing theoretical results with experimental quantities
it is usually better to 
use the charged pion mass since most experiments are performed with these.
When extracting quark masses, it is better to use the neutral pion mass
since this is expected to have only a very small contribution from
electromagnetism. In the below we use the values
\begin{equation}
M_{\pi^+} = 139.57018(35)~\mathrm{MeV}~,\qquad
M_{\pi^0} = 134.9766(6)~\mathrm{MeV}~.
\end{equation}
The pion decay constant is measured in $\pi\to\mu\nu$ and the main
uncertainty is the size of electromagnetic effects.
$V_{ud}$ is known to sufficient precision from neutron and nuclear decays.
We will adopt the value
\begin{equation}
\label{eq:Fpi}
F_\pi = 92.2\pm0.1~\mathrm{MeV}
\end{equation}
from the PDG \cite{Beringer:1900zz}.

The next major input needed are the $\pi\pi$ scattering lengths
and related quantities.
The main theoretical underpinning of this are the Roy equations
\cite{Roy:1971tc}, a set of integral equations that the $\pi\pi$ scattering
amplitude satisfies because of crossing and unitarity.
These require as input two subtraction constants, or equivalently
values of the scattering lengths $a_0^0$ and $a^2_0$, and
phenomenological input for the 
higher waves and at short distances. A large (re)analysis was done
in \cite{Ananthanarayan:2000ht}. This analysis confirmed a number of results
from the 1970s and sharpened them.
In \cite{Colangelo:2001df} the analysis was strengthened by two additional
inputs: The scalar radius should have a value of about
  $0.6$~fm$^2$ and 
the \chpt\ series for the $\pi\pi$ amplitude was found to converge
extremely well in the center of the Mandelstam triangle. This allowed to make
a rather sharp prediction for the two subtraction constants if one assumes
\chpt. The resulting values for the scattering lengths $a^0_0$ and $a^2_0$
have since been confirmed by other theoretical analyses
using the same or similar
methods~\cite{DescotesGenon:2001tn,GarciaMartin:2011cn}.
There has since also been a large experimental effort to pin down these
two scattering lengths by NA48/2 and others in both
$K_{\ell4}$ \cite{Batley:2007zz} and $K\to 3\pi$ \cite{Batley:2005ax}
decays and by DIRAC \cite{Adeva:2005pg}.
The values we will use for the scattering lengths are those from
 \cite{Colangelo:2001df}:
\begin{equation}
\label{eq:a0a2}
a^0_0 = 0.220\pm0.005\,,\qquad a^2_0= -0.0444\pm0.0010\,.
\end{equation}

The pion scalar form factor is not directly measureable but it can be obtained
from dispersion relations. The main part is given by the Omn\'es
solution using the $\pi\pi$ S-wave from the analysis above but improvements
are possible by including also other channels, first and foremost the
$KK$ channel. The main conclusion from
\cite{Donoghue:1990xh,Moussallam:1999aq} is that
\begin{eqnarray}
\label{eq:fs}
F_S^\pi(q^2) &=& F_S^\pi(0)\left(1+\frac{\langle
  r^2\rangle^\pi_S}{6}q^2+c^\pi_S q^4+\dots\right) 
\nonumber\\
\langle r^2\rangle^\pi_S &=& 0.61\pm0.04~\mathrm{fm}^2,
\qquad
c^\pi_S = 11\pm1~\mathrm{GeV}^{-4}~.
\end{eqnarray}
A more recent discussion can be found in \cite{Ananthanarayan:2004xy}.

The pion vector form factor can be measured directly as well as treated
with dispersive methods. The direct fit to the low-energy data,
which are dominated by \cite{Amendolia:1986wj}, was done in
\cite{Bijnens:1998fm}: 
\begin{eqnarray}
F_V^\pi(q^2) &=& F_V^\pi(0)\left(1+\frac{\langle
  r^2\rangle^\pi_V}{6}q^2+c^\pi_V q^4+\dots\right) 
\nonumber\\
\langle r^2\rangle^\pi_V &=& 0.437\pm0.016~\mathrm{fm}^2,
\qquad
c^\pi_V = 3.85\pm0.60~\mathrm{GeV}^{-4}~.
\end{eqnarray}
A more recent analysis using dispersive methods
\cite{Ananthanarayan:2013dpa} reached 
$\langle r^2\rangle^\pi_V\in (0.42,0.44)~\mathrm{fm}^2$ and
$c^\pi_V \in (3.79,4.00)~\mathrm{GeV}^{-4}$ in good agreement with the above
but somewhat smaller errors.

\subsection{Values of the LECs}
\label{nf2fits}

When looking at values of the LECs in two-flavour theory, the usual
convention is to use the $\bar l_i$ defined in (\ref{eq:deflib}).
These are independent of the subtraction scale but do depend  instead
explicitly on the pion mass. For a subtraction scale $\mu=0.77$~GeV
one should keep in mind that
\begin{equation}
-\ln\frac{M_\pi^2}{\mu^2} = 3.42\,.
\end{equation}
Values of the $\bar l_i$ around this value are thus dominated by the pion
chiral logarithm.

The values of $\bar l_1$ and $\bar l_2$ were originally determined
from the $D$-wave $\pi\pi$ scattering lengths in
\cite{Gasser:1983yg}. The analysis in \cite{Colangelo:2001df} relies instead
on the whole $\pi\pi$ scattering analysis and yields
\begin{equation}
\label{eq:l1l2}
\bar l_1 = -0.4\pm0.6\,,\qquad \bar l_2 = 4.3\pm0.1\,.
\end{equation}
The main sources of uncertainty are the input estimates of
the $c_i^r$ for both.

$\bar l_3$ is very difficult to get from phenomenology.
For this one needs to know how the pion mass depends on higher powers of
$\hat m$. Putting an upper limit on this leads to the estimate given
in \cite{Gasser:1983yg}:
\begin{equation}
\label{eq:l3}
\bar l_3 = 2.9\pm2.4\,.
\end{equation}

The scalar radius was used as input in \cite{Colangelo:2001df}. From the
value above they derived
\begin{equation}
\label{eq:l4}
\bar l_4 = 4.4\pm0.2\,.
\end{equation}
This value is in good agreement with the determination done in
\cite{Bijnens:1998fm} directly from the scalar radius.

The constant $\bar l_5$ is quite well known. It can be determined from the
difference between vector and axial-vector two-point functions.
The analysis in \cite{GonzalezAlonso:2008rf} quotes
\begin{equation}
\label{eq:l5}
\bar l_5 = 12.24\pm0.21\,.
\end{equation}

The constant $\bar l_6$ can be determined from the pion electromagnetic
radius. This was done in \cite{Bijnens:1998fm} and gives the value
\begin{equation}
\label{eq:l6}
\bar l_6 = 16.0\pm0.5\pm0.7\,.
\end{equation}
The last error is mainly from the estimate of the $c^r_i$.

A combination of the latter two can be obtained from the axial form factor
$F_A$
in the decay $\pi^+\to e^+\nu\gamma$. The two-loop calculation was done
in \cite{Bijnens:1996wm} with the result
\begin{equation}
\label{eq:l5l6}
\bar l_5-\bar l_6 = - 3.0\pm0.3\,.
\end{equation}
Ref.~\cite{Bijnens:1996wm} used $F_A= 0.0116\pm0.0016$ from the measured
value for $\gamma=F_A/F_V$ and the CVC prediction for $F_V$.
$F_V$ and $F_A$ have since been measured with better precision
with the result from \cite{Bychkov:2008ws} being
$F_A=0.0117\pm0.0017$ so the value in (\ref{eq:l5l6}) does not change.

Note that the values in (\ref{eq:l5}) and (\ref{eq:l5l6}) can be
combined to  \cite{GonzalezAlonso:2008rf}
\begin{equation}
\label{eq:l6b}
\bar l_6 = 15.24\pm0.39\,,
\end{equation}
nicely compatible within errors with  (\ref{eq:l6}),
thus providing proof that \chpt\ works
well in this sector. Another check that \chpt\ works was done
by looking at the relations for $\pi\pi$ scattering found in
\cite{Bijnens:2009zd} independent of the values of the $c_i^r$.
Those were fairly well satisfied.

The determination of the constants at order $p^6$ is in worse shape.
Four combinations of the $c_i^r$ are reasonably well known.
These come from $c_V^\pi$, $c_S^\pi$ \cite{Bijnens:1998fm} and
$\pi\pi$ scattering \cite{Colangelo:2001df}:
\begin{eqnarray}
\label{eq:ci} 
r_{V2}^r &=& -4 c_{51}^r+4 c_{53}^r = (1.6\pm0.5)\cdot 10^{-4}\,,
\nonumber\\
{r_{S3}^r} &=& {-8 c_6^r \approx  1.5  \cdot 10^{-4}}\,,
\nonumber\\
r_5^r &=& -8 c_1^r+10 c_2^r+14 c_3^r = (1.5\pm0.4)\cdot 10^{-4}\,,
\nonumber\\
r_6^r &=& 6 c_2^r+2c_3^r = (0.40\pm0.04)\cdot 10^{-4}\,.
\end{eqnarray}
$r_{S3}^r$ is not so well known since $c_S^\pi$ is dominated by the other
contributions.
 
\subsection{Including lattice results}
\label{sec:nf2lattice}

In the last years the quark masses obtainable on the lattice have been
coming closer to and even reaching the physical point. The situation
relevant for the 
quantities considered in this review is presented in the FLAG reports
\cite{Colangelo:2010et,Aoki:2013ldr}. In particular, Sec.~5.1 of
\cite{Aoki:2013ldr} reviews the status as of 2013. 

The value of $\bar l_6$ can be obtained from the lattice calculations
of the electromagnetic pion radius and $\bar l_4$ from the scalar radius.
These calculations have not yet reached a precision comparable to
(\ref{eq:l6}) and (\ref{eq:l4}). 
The combination $\bar l_1-\bar l_2$ can be obtained from higher-order
effects in the pion form factors. Again this value is not yet competitive
with the one of (\ref{eq:l1l2}).

In the continuum we cannot vary the pion mass but lattice QCD calculations
can easily do this by varying the quark masses. The constants that
influence this behaviour are thus much easier to obtain from lattice
calculations. The quantities that are measured here are
the variation of $F_\pi$ with the pion mass which gives $\bar l_4$
and the deviation of $M_\pi^2/(2B\hat m)$ from unity as a function of
$M_\pi$ which gives $\bar l_3$. The lattice calculations are done in a number
of physically different ways: with only up and down quarks, $N_f=2$,
including the strange quark, $N_f=2+1$, and including the charm quark as
well, $N_f=2+1+1$. In addition one often uses partially quenched
conditions where valence and sea quarks have different masses. 
The last case, $N_f=2+1+1$, is
pursued mainly by the ETM collaboration \cite{Baron:2011sf}.
The case with $N_f=2+1$ has many more contributors, 
RBC/UKQCD \cite{Arthur:2012opa}, MILC \cite{Bazavov:2010yq},
NPLQCD \cite{Beane:2011zm} and BMW \cite{Borsanyi:2012zv}.
For a more complete reference list, including  $N_f=2$, see Ref. 
\cite{Aoki:2013ldr}.
The FLAG results for $N_f=2$ and $N_f=2+1+1$
are dominated by the ETM results \cite{Baron:2009wt} and
\cite{Baron:2011sf}, respectively,
while the $N_f=2+1$ results are averaged over several collaborations,
which are not always in good agreement.
This is the reason why the errors for the $N_f=2+1$ case as quoted
below are larger.
The results quoted are always those when the various $N_f$ cases have been
analysed with \nftwo\ \chpt. FLAG \cite{Aoki:2013ldr} gives the averages
\begin{eqnarray}
\label{eq:flagsu2}
\left.\bar l_3\right|_{N_f=2} &=& 3.45\pm0.26\,,\quad
\left.\bar l_3\right|_{N_f=2+1} = 2.77\pm1.27\,,\quad
\left.\bar l_3\right|_{N_f=2+1+1} = 3.70\pm0.27\,,
\nonumber \\
\left.\bar l_4\right|_{N_f=2} &=& 4.59\pm0.26\,,
\left.\bar l_4\right|_{N_f=2+1} = 3.95\pm0.35\,,\quad
\left.\bar l_4\right|_{N_f=2+1+1} = 4.67\pm0.10\,.
\nonumber \\
 & &
\end{eqnarray}
Clearly, there is still a significant spread in central
  values. Not included in the FLAG averages (\ref{eq:flagsu2}) are the
  more recent $N_f=2+1$ results \cite{Durr:2013goa} $\bar l_3 = 2.5\pm
  0.7$ and $\bar l_4 = 3.8\pm0.5$.

A rough estimate
that covers the lattice range and the continuum results 
(\ref{eq:l3},\ref{eq:l4}) is
\begin{equation}
\label{eq:l3l4}
\bar l_3 = 3.0\pm0.8\,,\quad
\bar l_4 = 4.3\pm0.3\,.
\end{equation}
These are what we will use in the fits for the \nfthree\ constants below.

\section{THREE FLAVOURS}
\label{sec:nf3}

\subsection{New issues}

Among the issues specific to the three-flavour case are the 
size of NNLO corrections of $O(p^6)$ or alternatively
the convergence of the low-energy expansion for chiral
$SU(3)$. Related to this issue is a possible paramagnetic 
effect \cite{DescotesGenon:1999uh}, which would manifest itself here by
rather large values for $L_4^r$ and $L_6^r$. The values we obtain
are not fully conclusive but large $N_c$, Sec.~\ref{sec:largenc},
and the present lattice results, Sec.~\ref{sec:nf3lattice}, 
  support small values for $L_4^r$ and $L_6^r$.

The convergence of several physical quantities is discussed in
Sec.~\ref{sec:convergence}.

\subsection{Large $N_c$}
\label{sec:largenc}

The expansion in the number of colours was defined in \cite{'tHooft:1973jz}.
Leaving aside $l_7$ and $L_7$ because of the special large-$N_c$
counting due to $\eta^\prime$ exchange \cite{Gasser:1984gg,Peris:1994dh}, 
there is an important difference between the LECs of $O(p^4)$ for
chiral $SU(2)$ and $SU(3)$.

The $SU(2)$ LECs $l_1, \dots, l_6$ are all
leading order in $1/N_c$, i.e. of $O(N_c)$. In the $SU(3)$ case, there
are three (combinations of) LECs that are suppressed and of $O(1)$ at
large $N_c$: $2 L_1 - L_2$, $L_4$ and $L_6$ \cite{Gasser:1984gg}. It
is of special interest whether the phenomenological analysis respects
this hierarchy.

In the original analysis of Gasser and Leutwyler \cite{Gasser:1984gg}
the large-$N_c$ suppressed LECs were assumed to vanish at a scale
$\mu=M_\eta$. More recent analyses showed (e.g., fit All in
Ref.~\cite{Bijnens:2011tb}) that it is difficult to verify the
large-$N_c$ suppression with global fits especially for $L_4$ 
and $L_6$ because of big errors.

Lattice determinations of $L_4$ and $L_6$, on the other hand, are
quite consistent with the large-$N_c$ suppression
\cite{Aoki:2013ldr}. Why is it then notoriously difficult to
extract meaningful values for $L_4$ and $L_6$ from global fits?

In the case of $L_4$, a partial explanation is the apparent
anti-correlation of $L_4$ with the leading-order LEC $F_0$ in the fits of
Ref.~\cite{Bijnens:2011tb}: the bigger $F_0$, the smaller
$L_4^r(M_\rho)$, and vice versa.
This anti-correlation can be understood to some extent from the
structure of the chiral $SU(3)$ Lagrangian up to and including NLO:  
\noindent
\begin{eqnarray}
\label{eq:anticorr}
\mathcal{L}_2  +  \mathcal{L}_4^{\nfthree} &=&
\displaystyle\frac{F_0^2}{4} \langle u_\mu u^\mu + \chi_+
 \rangle
+ ~L_4 \langle  u_\mu u^\mu \rangle 
\langle  \chi_+ \rangle + \dots  \nn
&&  \hspace*{-2cm} = ~\displaystyle\frac{1}{4} \langle  u_\mu u^\mu
 \rangle \left[ F_0^2 + 8 L_4 \left(2 M^{\circ 2}_K +
  M^{\circ 2}_{\pi} \right)\right] + \dots~ 
\end{eqnarray} 
where $M^\circ_P (P=\pi,K)$ denotes the lowest-order meson masses. The
dots refer to the remainder of the NLO Lagrangian (\ref{eq:NLOlag3})
in the first
line and to terms of higher order in the meson fields in the second
line. Therefore, a lowest-order tree-level contribution is always
accompanied by an $L_4$ contribution in the combination
\cite{Ecker:2013pba} 
\begin{equation} 
\label{eq:Fmu}
F(\mu)^2:=F_0^2 + 8 L_4^r(\mu) \left(2 M^{\circ 2}_K +
 M^{\circ 2}_{\pi} \right)~.
\end{equation}
Of course, there will in general be additional contributions involving
$L_4$ at NLO, especially in higher-point functions. Nevertheless, the
observed anti-correlation between 
$F_0$ and $L_4$ is clearly related to the structure of the chiral
Lagrangian. Note that $F^2_{\pi}/16 M_K^2 = 2 \times 10^{-3}$ is the
typical size of an NLO LEC. Although of different chiral order, the two
terms in $F(\mu)^2$ could a priori be of the same order of magnitude.
This makes it very difficult to disentangle $F_0$ and $L_4$ in
phenomenological fits. At least in principle, the lattice is better
off in this respect because the masses of quarks and mesons can be
tuned on the lattice. In practice, most lattice studies employ strange
quark masses of similar size corresponding to the actual kaon mass. As
Eq.~(\ref{eq:Fmu}) indicates, the tuning of the light quark mass and
thus of the pion mass is less effective for disentangling $L_4$ and
$F_0$. Nevertheless, the
uncertainties of $L_4$ from lattice evaluations are definitely smaller
\cite{Aoki:2013ldr,Ecker:2013pba} than from continuum fits. Note also
that the combination $F(\mu)$ defined in (\ref{eq:Fmu}) can be much
better determined than $F_0$ itself \cite{Ecker:2013pba}.
A similar discussion can clearly be done for $L_6^r$ and 
the lowest-order mass term ${(F_0^2/4)}\trf{\chi_+}$.

In the following analysis, we are therefore not going to extract the
large-$N_c$ suppressed LECs directly from the global fits. However, it
will turn out to be sufficient to restrict $L_4^r$ to a
reasonable range suggested by large $N_c$ and lattice results. The
LECs $L_6$ and $2 L_1 - L_2$ will then more or less automatically
follow suit.

\subsection{Continuum data}
\label{sec:input3}

All \chpt\ results we use are known to NNLO and since long at NLO.
The results for the masses and decay constants at NLO are from
\cite{Gasser:1984gg} and at NNLO from \cite{Amoros:1999dp}.
The scalar form factor of the pion was done at NLO in \cite{Gasser:1984ux}
and at NNLO in \cite{Bijnens:2003xg}. $\pi\pi$ scattering was done at NLO in
\cite{Gasser:1984gg} and at NNLO in \cite{Bijnens:2004eu}. $\pi K$
scattering was done at NLO in \cite{Bernard:1990kw} and at NNLO in
\cite{Bijnens:2004bu}. Finally, the $F$ and $G$ form factors in
$K_{\ell4}$ were done at NLO in \cite{Bijnens:1989mr,Riggenbach:1990zp}
and NNLO in \cite{Amoros:2000mc}.

We will use as input the values of $F_\pi$, $M_\pi$, $\langle r^2\rangle_S^\pi$,
$c_S^\pi$, $a^0_0$ and $a^2_0$
as given in Sec.~\ref{sec:input2}.

The remaining input has changed somewhat from \cite{Bijnens:2011tb} but
the final effect of these changes on the fits discussed below is small.
The main differences come from our different treatment of the $C_i^r$.

The value of $F_K/F_\pi$ we take
from the PDG \cite{Beringer:1900zz}:
\begin{equation}
\frac{F_K}{F_\pi} = 1.198\pm0.006\,.
\end{equation}
The error is dominated by the uncertainty of $V_{us}$. This value is
in good agreement with the lattice determinations \cite{Aoki:2013ldr}.

We also include the results on $\pi K$ scattering from a Roy-Steiner
analysis \cite{Buettiker:2003pp}:
\begin{equation}
\label{eq:RS}
a_0^{1/2} = 0.224 \pm 0.022\,,\quad a^{3/2}_0 = -0.0448 \pm 0.0077\,.
\end{equation}

The main change w.r.t. \cite{Bijnens:2011tb} is that we now
use the normalization of the $K_{\ell4}$ decay also from NA48/2.
The summary of their results can be found in \cite{Batley:2012rf}.
{}From those results we use for the two form factors $F$ and $G$
their slope and value at threshold defined as  
\begin{eqnarray}
\label{eq:Kl4}
F &=& f_s + f^\prime_s q^2+\dots\,, \quad f_s=5.705\pm0.035\,,
\quad f^\prime_s = 0.867\pm0.050\,,
\nonumber\\
G &=& g_p+ g^\prime_p q^2+\dots\,,\quad g_p = 4.952\pm0.086\,,
\quad g^\prime_p = 0.508\pm0.122\,,
\end{eqnarray}
with $q^2 = s_{\pi\pi}/(4m_\pi^2)-1$.

The final input we need is the quark mass ratio. We perform fits for several
values of these but use as central value \cite{Aoki:2013ldr}
\begin{equation}
\label{eq:msmhat}
\frac{m_s}{\hat m} = 27.5\pm0.5\,.
\end{equation}

The masses used are those from the PDG (with the PDG 2010 value for $M_\eta$,
since rerunning all the input calculations would be very time consuming;
this small difference is not of any relevance in the remainder):
\begin{equation}
M_{K^+} = 493.677(16)~\mathrm{MeV},\quad
M_{K^0} = 497.614(24)~\mathrm{MeV},\quad
M_\eta = 547.853(18)~\mathrm{MeV}\,.
\end{equation}
When using the masses for the quark mass ratios and decay constants
we use the neutral pion mass, the eta mass and the average kaon mass
with the electromagnetic corrections removed using the estimate of
\cite{Bijnens:1996kk}. This
results in an average kaon mass of
\begin{equation}
M_K = 494.5~\mathrm{MeV}\,.
\end{equation}

\subsection{Continuum fits}
\label{sec:nf3fits}

The main principle of the fit is the same as in \cite{Bijnens:2011tb}.
We calculate numerically the $p^4$ and $p^6$ corrections for all the
quantities discussed above. In all cases we use the physical masses
and $F_\pi$ in the expressions.
For $F_K/F_\pi$ we use the form
\begin{equation}
\frac{F_K}{F_\pi} = 1+F_K^{(4)}-F_\pi^{(4)}+
\left[F_K^{(6)}-F_\pi^{(6)}-F_\pi^{(4)}\left(F_K^{(4)}-F_\pi^{(4)}\right)\right]\,.
\end{equation}
The masses are included via $M^2_M = M_{M0}^2+M_M^{2(4)}+M_M^{2(6)}$
and we then add as a $\chi^2$ that $m_s/\hat m$
obtained from $M^2_{K0}/M^2_{\pi0}$ and $M^2_{\eta0}/M^2_{\pi0}$ 
should agree with (\ref{eq:msmhat}) with an error of 5\%.
This was chosen as a reasonable compromise for the neglected
higher-order terms.
The exception to the errors quoted in Secs. \ref{sec:input2}
and \ref{sec:input3} is that we double the errors for $a^{1/2}_0$ and
$a^{3/2}_0$ in (\ref{eq:RS}). 

The fits are also done with the variable
\begin{equation}
L_A^r = 2L_1^r-L_2^r
\end{equation}
to allow for a large-$N_c$ test. 

\subsubsection{NLO fits}
\label{sec:NLOfits}

We first perform a number of fits at NLO with $F_K/F_\pi$,
$m_s/\hat m$ from the masses $M_K$ and $M_\eta$, the four scattering
lengths, the pion scalar radius 
and the $K_{\ell4}$ form factors of (\ref{eq:Kl4}) as input. We thus
have 12 inputs to determine the $L_i^r$ for $i=1,\dots,8$.

The results are shown in Table~\ref{tab:p4fits}. The free fit in the
second column has the smallest $\chi^2$ but it does not exhibit the
large-$N_c$ hierarchy of the LECs. Moreover, the large values of
$L_4^r$ and $L_6^r$ are in conflict with most lattice results
\cite{Aoki:2013ldr}. The anti-correlation with 
$F_0$ discussed in Sec.~\ref{sec:largenc} is manifest. We also note
that the results of this fit and the last column of Table~5 in
\cite{Bijnens:2011tb} are a little different but compatible. The
  small differences are due to the different errors that have been
  used here, rather than the small change in central values. 

As emphasized in Sec.~\ref{sec:largenc}, the lattice is in a much
better position to determine $L_4$. We have therefore
restricted $L_4^r$ to a range compatible with lattice
studies. In columns 3, 4, 5 in Table \ref{tab:p4fits}, we display the
results for $10^3 L_4^r = 0,\, 0.3,\, - 0.3$, respectively. What is
remarkable is the manifest positive correlation of $L_4$ with $L_A$
and $L_6$: Enforcing large $N_c$ on  $L_4^r$ makes also $L_A^r$ and
$L_6^r$ small, in agreement with the large-$N_c$ suppression.

The relatively large $\chi^2$ in the restricted fits comes almost
exclusively from the scattering lengths $a^2_0$ and $a^0_0$. These
parameters were determined in a sophisticated dispersion theoretical
analysis of pion-pion scattering \cite{Colangelo:2001df} in the
framework of chiral $SU(2)$. It is therefore not surprising that an
NLO fit in chiral $SU(3)$ cannot really cope with the precision of 
$a^2_0$ and $a^0_0$ in (\ref{eq:a0a2}). However, doubling the errors of
$a^2_0$, $a^0_0$ does not really change the picture: The resulting fit
values for the $L_i^r$ (including errors) are almost unchanged, but
the $\chi^2$ decreases to values similar to the one of the free fit in
column 2. 

Therefore, we consider the combined fit values with restricted $L_4^r$
in column 6 as the most realistic values of the $L_i^r$ at NLO. The
errors listed are based on the input errors only and do not reflect
the uncertainties due to higher-order corrections. This statement does
not apply to the original estimates of the $L_i$ by Gasser and
Leutwyler \cite{Gasser:1984gg} reproduced in the last column for
comparison. It is worth emphasizing that our $p^4$ fit values and the
estimates in Ref.~\cite{Gasser:1984gg} from nearly 30 years ago (columns
6 and 7) are still compatible with each other. 

There are two more LECs, $L_9^r$ and $L_{10}^r$.
These were determined via the pion electromagnetic radius and the
axial form factor in $\pi\to e\nu\gamma$ in
\cite{Gasser:1984gg} with the results
\begin{equation}
L_9^r = (6.9\pm0.7) \cdot 10^{-3}\,,\qquad
L_{10}^r = (-5.5\pm0.7)\cdot 10^{-3}\,.
\end{equation}
The value for $L_9^r$ was redone at NLO in \cite{Bijnens:2002hp}
with essentially the same result. A more precise value for $L_{10}^r$
was obtained in \cite{GonzalezAlonso:2008rf} as
\begin{equation}
L_{10}^r = (-5.22\pm0.06)\cdot 10^{-3}\,.
\end{equation}

\begin{table}[htb!]
\caption{\label{tab:p4fits} Fits at NLO
 with $L_4^r$ free, $0, \, 0.3 \cdot 10^{-3}$ and $- 0.3 \cdot
 10^{-3}$, respectively. The last two columns contain the combined
 results for the range $-0.3 \cdot 10^{-3} \le L_4^r \le 0.3 \cdot
 10^{-3}$ and for comparison the original values from
 Ref.~\cite{Gasser:1984gg}. Except for the last column
 \cite{Gasser:1984gg}, the errors are only due to the input errors,
 no estimate of the error due to higher orders is included. As always
 for $SU(3)$ LECs, the renormalization scale is $\mu=0.77$ GeV.}
\begin{center}
\begin{tabular}{@{}crrrrrr@{}}
\hline\hline
& & & & & &\\[-.3cm] 
 & free fit & & & & $p^4$ fit & Ref.~\cite{Gasser:1984gg} \\[.1cm]  
\hline \\[-.3cm] 
$10^3 L_A^r$&   1.17(27)&   0.39(16)&   0.52(16)&   0.27(16)&
0.4(2)&   \\ 
$10^3 L_1^r$&   1.11(10)&   0.98(09)&   1.00(09)&   0.95(09)&
1.0(1)&   0.7(3)\\ 
$10^3 L_2^r$&   1.05(17)&   1.56(09)&   1.48(09)&   1.64(09)&
1.6(2)&   1.3(7) \\ 
$10^3 L_3^r$&$-$3.82(30)&$-$3.82(30)&$-$3.82(30)&$-$3.82(30)&
$-$3.8(3)& $-$4.4(2.5) \\     
$10^3 L_4^r$&   1.87(53)&$\equiv$0  &$\equiv$0.3&$\equiv-$0.3&
0.0(3)  &$-$0.3(5) \\ 
$10^3 L_5^r$&   1.22(06)&   1.23(06)&   1.23(06)&   1.23(06)&
1.2(1)&   1.4(5)\\ 
$10^3 L_6^r$&   1.46(46)&$-$0.11(05)&   0.14(06)&$-$0.36(05)&
0.0(4)&  $-$0.2(3)\\ 
$10^3L_7^r$&$-$0.39(08)&$-$0.24(15)&$-$0.27(14)&$-$0.21(17)&
$-$0.3(2)  &$-$0.4(2)\\  
$10^3 L_8^r$&   0.65(07)&   0.53(13)&   0.55(12)&   0.50(14)&
0.5(2)&   0.9(3)\\ 
\hline \\[-.3cm]                                                        
$\chi^2$    &   3.8     & 16        & 12        & 20        & 
&       \\ 
dof         &   4       &  5        &  5        &  5        &   5
&        \\ 
$F_0$ [MeV] &   58      &  81       &  76       &  86       &   81(5)
&      \\ 
\hline
\end{tabular}
\end{center}
\end{table}

\subsubsection{NNLO fits}
\label{sec:NNLOfits}

We now add more input. In addition to $c_S^\pi$ defined in
(\ref{eq:fs}), we include also the $\bar l_i$ discussed in
Sec.~\ref{nf2fits}. The latter can be calculated from the $L_i$ 
at $O(p^6)$ using the results of Ref.~\cite{Gasser:2007sg}.  As was 
already remarked in that reference, this will allow a handle on some
of the large-$N_c$ suppressed $C_i^r$. In addition, we require 
a not too badly behaved perturbative \chpt\ series for the
masses. This we enforce by adding to the $\chi^2$ 
a contribution of the type $f^\chi(M^{2(6)}_M/M_M^2/\Delta_M)$ with
\begin{equation}
f^\chi(x) = 2x^4/(1+x^2)
\end{equation}
and $\Delta_M=0.1$.
This form was chosen to be one when x=1, quadratic for large $x$ but
turning on slower than $x^2$. If we had chosen $f^\chi(x)=x^2$ it would be
like a normal $\chi^2$.

NNLO fits of the $L_i$ turn out to be very sensitive to the values of
the $C_i$. The naive dimensional estimate\footnote{We shall always
  display the $C_i$ in units of $10^{-6}$.} for 
the NNLO LECs is $C_0 = (4 \pi)^{-4} = 40 \cdot 10^{-6}$.

To set the scene, we perform NNLO fits for two different scenarios. In
the first case, all $C_i$ are set to zero at the scale $\mu=0.77$
GeV. For the second scenario, we take the predictions of a chiral
quark model \cite{Jiang:2009uf}, mainly because it is the only model
we are aware of that predicts all the $C_i$ contributing to the
observables in our fits. The results are displayed in Table
\ref{tab:Ca0chinafits}. In both cases, the fit is not satisfactory:
In addition to the large $\chi^2$, the LECs $L_A$, $L_4$ and $L_6$ show
no sign of large-$N_c$ suppression.

\begin{table}[tb!]
\caption{\label{tab:Ca0chinafits} Fits at NNLO for two different
  choices of the $C_i$. The results in the second column were obtained
  by taking $C_i^r=0 ~(\forall i)$, those in the third column are based
  on the $C_i$ from a chiral quark model \cite{Jiang:2009uf}. The 
  renormalization scale is $\mu=0.77$ GeV.}
\begin{center}
\begin{tabular}{@{}crr@{}}
\hline\hline
& & \\[-.3cm] 
$C_i$ & \hspace*{.7cm} $C_i^r=0$ & \hspace*{.7cm}
Ref.~\cite{Jiang:2009uf} \\ 
\hline \\[-.3cm] 
$10^3 L_A^r$&   1.17(12) &  0.70(12) \\
$10^3 L_1^r$&   0.67(06)&   0.48(07) \\
$10^3 L_2^r$&   0.17(04)&   0.25(04) \\
$10^3 L_3^r$&$-$1.76(21)& $-$1.68(22)  \\
$10^3 L_4^r$&   0.73(10)& 0.86(11)   \\
$10^3 L_5^r$&   0.65(05)&   2.08(14)  \\
$10^3 L_6^r$&   0.25(09)&  0.83(06)  \\
$10^3L_7^r$&$-$0.17(06)&$-$0.33(06)  \\
$10^3 L_8^r$&   0.22(08)&  1.03(14)  \\
\hline \\[-.3cm]                                                        
$\chi^2$    &   26     & 41   \\
dof         &   9       &  9   \\
\hline
\end{tabular}
\end{center}
\end{table}

Therefore, it is obvious that we have to make some assumptions about
the NNLO LECs in order to proceed. There are altogether 34
(combinations of the) $C_i$ that appear in our 17 input observables.  
For most of those $C_i$ predictions are available in the literature,
although in some cases contradictory. There are essentially three
types of predictions. The estimates in the first group are mainly
phenomenologically oriented
\cite{Durr:1999dp,Boito:2013qea,Bijnens:2003uy,Bijnens:2003xg,
Jamin:2004re,Kampf:2006bn,Prades:2007ud,Unterdorfer:2008zz,
GonzalezAlonso:2008rf,Masjuan:2008fr,Bernard:2009ds,Boito:2012nt}. 
A  second class uses more theory in addition to phenomenological input, 
e.g., chiral quark models, resonance chiral theory, short-distance
constraints, holography, etc. 
\cite{Knecht:2001xc,Cirigliano:2004ue,Cirigliano:2005xn,Rosell:2006dt,
Cirigliano:2006hb,Kaiser:2007zz,Pich:2008jm,Jiang:2009uf,
SanzCillero:2009ap,Pich:2010sm,Colangelo:2012ipa,Golterman:2014nua}.  
Finally, there are also some estimates from lattice studies 
\cite{Bazavov:2010hj,Ecker:2010nc,Bazavov:2012cd,Boyle:2014pja,Ecker:2013pba}.  

We use the available information to define priors for the $C_i$ with
associated ranges of acceptable values. The fits are then performed by
two different methods leading essentially to the same results:
minimization and a random-walk procedure in the restricted space of
the $C_i$. If the resulting fit values for the $L_i$ deviate too much
from the $p^4$ values in Table \ref{tab:p4fits} and/or if the $\chi^2$
is too large, we modify the boundaries of the $C_i$ space and start
again. 

Therefore, we cannot claim to have found the best values for the
$L_i$ to NNLO with this procedure, because the notion of ``best
values'' is mathematically ill-defined with 17 input data and
8+34 parameters. On the other hand, our final results displayed in
Table \ref{tab:p6fits} exhibit the following attractive properties.

\begin{itemize} 
\item As shown by the $\chi^2$, especially in comparison with the test
  fits in Table \ref{tab:Ca0chinafits}, the quality of the fits is
  excellent.
\item The values of the NLO and NNLO fits are of course different, but not
  drastically so. This has partly been enforced by requiring that the
  meson masses show a reasonable ``convergence'' (see
  Sec.~\ref{sec:convergence}).
\item The results to NNLO are much more sensitive to $L_4$ than at
  NLO. We therefore present 
  only two cases, one without restricting $L_4^r$  and
  the other for fixed $L_4^r=0.3 \cdot 10^{-3}$. In the latter case, as
  found already at NLO, the small $L_4$ guarantees that also 
  $L_A$ and $L_6$ are in accordance with large $N_c$.
\end{itemize} 

In Table \ref{tab:p6fits} we present both cases: our preferred fit
(BE14) with $L_4^r=0.3 \cdot 10^{-3}$ and the general fit without any
restrictions on the $L_i^r$. It is clear that our preference is not
based on $\chi^2$ only, but on a good deal of theoretical prejudice as
well. Moreover, our preferred fit BE14 
must always be considered together with the values of the NNLO LECs
collected in Table \ref{tab:Cafit}. We do not display the $C_i$ for
the unrestricted fit but the values are similar.   

\begin{table}[tb!]
\caption{\label{tab:p6fits} NNLO fits for the LECs $L_i^r$. The second
  column contains our preferred fit (BE14) with fixed $L_4^r= 0.3
  \cdot 10^{-3}$, the third one the general free fit without any
  restrictions on the $L_i^r$. No estimate of the error due to higher
  orders is included.} 
\begin{center}
\begin{tabular}{@{}crr@{}}
\hline\hline 
fit  & \hspace*{.7cm} BE14 & \hspace*{.7cm} free fit \\
\hline
$10^3 L_A^r$&   0.24(11)&  0.68(11) \\
$10^3 L_1^r$&   0.53(06)&    0.64(06) \\
$10^3 L_2^r$&   0.81(04)&  0.59(04) \\
$10^3 L_3^r$&$-$3.07(20)&$-$2.80(20)\\
$10^3 L_4^r$&$\equiv$0.3&  0.76(18)  \\
$10^3 L_5^r$&   1.01(06)&  0.50(07) \\
$10^3 L_6^r$&   0.14(05)&  0.49(25) \\
$10^3 L_7^r$&$-$0.34(09)&$-$0.19(08)\\
$10^3 L_8^r$&   0.47(10)&  0.17(11) \\
\hline                                                        
$\chi^2$    &   1.0     & 0.5        \\
$F_0$ [MeV] &   71      &  64
       \\
\hline
\end{tabular}
\end{center}
\end{table}

\begin{table}[tb!]
\caption{\label{tab:Cafit} Preferred values of the NNLO LECs for
  fit BE14 in Table \ref{tab:p6fits}. As always, the renormalization
  scale is $\mu=0.77$ GeV and the numerical values are in units of
  $10^{-6}$.} 
\begin{center}
\begin{tabular}{@{}cr@{\hspace*{7mm}}cr@{\hspace*{7mm}}cr@{\hspace*{7mm}}cr@{}}
\hline\hline
LEC  &  & LEC & & LEC & & LEC & \\
\hline
$C_1$  &  12     &$C_{11}$& $-$4.0& $C_{20}$ & 1.0     &$C_{29}$  & $-$ 20  \\
$C_2$  &  3.0    &$C_{12}$& $-$2.8& $C_{21}$ & $-$ 0.48&  $C_{30}$ & 3.0    \\
$C_3$  &  4.0    &$C_{13}$&   1.5 & $C_{22}$ & 9.0     & $C_{31}$  & 2.0    \\
$C_4$  &  15     &$C_{14}$& $-$1.0&  $C_{23}$ & $-$ 1.0 &$C_{32}$   & 1.7   \\
$C_5$  & $-$4.0  &$C_{15}$& $-$3.0& $C_{25}$   & $-$ 11  & $C_{33}$ & 0.82  \\
$C_6$  & $-$4.0  &$C_{16}$&   3.2 &   $C_{26}$ & 10    &   $C_{34}$  & 7.0 \\
$C_7$  &  5.0    &$C_{17}$& $-$1.0& $C_{28}$   & $-$ 2.0 & $C_{36}$  & 2.0 \\
$C_8$  &  19     &$C_{18}$&   0.63& \multicolumn{3}{l}{$C_{63} - C_{83} + C_{88}/2 $}&  $-$ 9.6 \\ 

$C_{10}$&  $-$0.25&$C_{19}$ & $-$ 4.0 & \multicolumn{3}{l}{$C_{66} - C_{69} - C_{88} + C_{90}$} & 50 \\
\hline
\end{tabular}
\end{center}
\end{table}

There have been some more studies of $L_4^r$, $L_6^r$ and the $C_i^r$
using the pion and kaon scalar form factor \cite{Bijnens:2003xg},
$\pi\pi$ scattering \cite{Bijnens:2004eu} and
$\pi K$ scattering \cite{Bijnens:2004bu,Kampf:2006bn}. Most of these studies
were very sensitive to the input values of the $L_i^r$ assumed and the input
data used are to a large extent included in the fits discussed
above. We therefore do not discuss those constraints.
The constraints from the scalar $K \pi$
transition form factor, or $f_0$ in $K_{l3}$ decays can in principle also be
used. The \chpt\ result is in \cite{Bijnens:2003uy} and has been
used in \cite{Jamin:2004re,Bernard:2007tk} to obtain constraints on $C_{12}^r$
and $C_{34}^r$ but the results depend again on the $L_i^r$ input used.

The situation for the constants $L_9^r$, $L_{10}^r$ and associated $C_i^r$
is better.
The pion electromagnetic radius is dominated by the $L_9^r$ contribution.
The determination in \cite{Bijnens:2002hp} to NNLO gives
\begin{equation}
\label{eq:8890}
L_{9}^r = (5.93\pm0.43)\cdot 10^{-3}\,,
\qquad
C_{88}^r-C_{90}^r = (-55\pm5)\cdot 10^{-6}\,.
\end{equation}
The values of $L_{10}^r$ and $C_{87}^r$ can be obtained from 
sum rules for the difference between the vector and axial-vector current
correlators. The required \chpt\ calculation was done in
\cite{Amoros:1999dp} and recent analyses of the spectral sum rules are
\cite{GonzalezAlonso:2008rf,Boito:2012nt,Boyle:2014pja}. The latter also use
some lattice data. A reasonable average of the values for $10^3 L_{10}^r$
of $-4.06\pm0.39$ \cite{GonzalezAlonso:2008rf}, $-3.1\pm0.8$ \cite{Boito:2012nt}
and ${ -3.46\pm0.32}$ \cite{Boyle:2014pja} is
\begin{equation}
L_{10}^r = (-3.8\pm0.4)\cdot 10^{-3}\,.
\end{equation}
The NNLO \chpt\ calculation for $\pi\to l\nu\gamma$ was done
\cite{Geng:2003mt} but has not been used to obtain a value for $L_{10}^r$.
The values for $C_{87}^r$ in the above references are compatible and give
\begin{equation}
\label{eq:C87}
C_{87}^r = (42\pm2)\cdot 10^{-6}\,.
\end{equation}
The error in (\ref{eq:C87}) is probably a little underestimated. It does
not include higher-order \chpt\ effects.
The large-$N_c$ estimate of \cite{Masjuan:2008fr},
$10^6C_{87}^r= 48\pm5$, is quite compatible with the above.

A similar analysis in the scalar sector for the difference
between scalar and pseudo-scalar spectral functions can in principle be done.
However, here one has to rely on much more theoretical input since direct
data are not available. Bounds on $L_6^r$ were derived in
\cite{Moussallam:1999aq,Moussallam:2000zf} with values typically a
little larger 
than those of the fits reported here. Results for $L_8^r$ were derived in
\cite{Bordes:2012ud,Rosell:2006dt} with values for
$10^3 L_8^r$ of $1.0\pm0.3$ and $0.6\pm0.4$, which are again somewhat
larger than our estimates. However, neither of these references included
a dependence on the other $L_i^r$ used as input.

\subsection{Including lattice results}
\label{sec:nf3lattice}

There have been a few papers combining continuum and lattice input
\cite{Bernard:2009ds,Ecker:2010nc,Ecker:2013pba,Boyle:2014pja}, but so
far no major 
effort has been done to combine the two in a systematic fashion.
The situation on the lattice side has been reviewed in the FLAG reports
\cite{Colangelo:2010et,Aoki:2013ldr}.

One of the problems is that relatively few lattice collaborations
actually use the full NNLO formulas to fit the data. Given that a typical
\nfthree \,\,\chpt\ correction at NLO is about 25\% and the expected NNLO
correction is thus about 7\%, it is clear that NLO \chpt\ will not be
sufficient to analyse lattice data at the physical kaon and pion
masses. On the other hand, using the same argument, a typical 
N$^3$LO correction would be of the order of 1.5\%, which is more
appropriate. 
The MILC collaboration \cite{Bazavov:2009fk,Bazavov:2010hj,Bazavov:2011fh}
is here one of the exceptions. The formulas are known
also at NNLO for all needed partially quenched cases
\cite{Bijnens:2004hk,Bijnens:2005ae,Bijnens:2005pa,Bijnens:2006jv}.
It should be noted that the number of new parameters is not that large.
Most lattice calculations use an analytic NNLO mass term in their fits.

We only quote here the results of MILC \cite{Bazavov:2009fk} and
HPQCD \cite{Dowdall:2013rya}. The former are with a full NNLO \chpt\ analysis
while the latter are with an NLO \chpt\ analysis augmented with analytic NNLO
terms. It is therefore not clear whether the latter should be compared
fully with our results. The results are given in Table
\ref{tab:lattice}.

\begin{table}[tb!]
\caption{\label{tab:lattice} Lattice results from the two most
  complete analyses available. The lattice values at $\mu=M_\eta$ have
been transformed to the usual scale $\mu=0.77$ GeV.}
\begin{center}
\begin{tabular}{@{}crrrr@{}}
\\[-.2cm] 
\hline\hline 
  & \cite{Bazavov:2009fk} & \cite{Dowdall:2013rya}\\
\hline
$10^3 L_4^r$ & 0.04(14) & 0.09(34)\\
$10^3 L_5^r$ & 0.84(40)& 1.19(25)\\
$10^3 L_6^r$ & 0.07(11) & 0.16(20)\\
$10^3 L_8^r$ & 0.36(09)& 0.55(15)\\
\hline
\end{tabular}
\end{center}
\end{table}

The results shown clearly live up to the large-$N_c$ expectations that
$L_4^r$ and $L_6^r$ should be small at $\mu=0.77$ GeV. The values for
$L_5^r$ and $L_8^r$ are compatible with the continuum estimates with small 
$L_4^r$ enforced.

The lattice results for the other $L_i^r$ have not yet reached the
accuracy needed to compete with the continuum determinations.

\subsection{Resonance saturation}
\label{sec:resonance}

In \chpt, LECs parametrize physics at shorter distances. Following the
time-honoured notion of vector meson dominance, Gasser and Leutwyler
\cite{Gasser:1983yg} suggested that the  $\rho$ meson should play a
special role in the LECs to which it contributes. This suggestion was
later formulated in terms of a resonance Lagrangian for chiral
$SU(3)$ \cite{Ecker:1988te,Ecker:1989yg}. Saturating the NLO LECs
$L_i$ with the lowest-lying resonance nonets turned out to provide a
qualitative understanding of the numerical values of the $L_i^r$ for
renormalization scales near $0.77$ GeV.

In this subsection, we first review the status of resonance saturation
in view of our fit results for the $L_i^r$, both at NLO and at
NNLO. In addition, we then investigate the validity of resonance
saturation also for some of the NNLO LECs $C_i^r$, which come with our
preferred fit BE14 of the $L_i^r$ (see Tables
\ref{tab:p6fits},\ref{tab:Cafit}).  

The expressions for the $L_i$ in terms of the parameters of the
lowest-lying vector, axial-vector and scalar multiplets (neglecting
small contributions from pseudoscalar resonances) are reproduced
in Table \ref{tab:p4res}. The resonance couplings were introduced in
Ref.~\cite{Ecker:1988te}. For the numerical estimates, we use the
well-known relations from short-distance constraints on spectral
functions and form factors 
\cite{Weinberg:1967kj,Knecht:1997ts,Ecker:1989yg,Ecker:1988te,Jamin:2001zq,Pich:2002xy}:  
\begin{eqnarray} 
\label{eq:reso}
F_V G_V = F_0^2~, & \qquad & 4 c_d c_m = F_0^2~, \no \\[.1cm]
F_V^2 - F_A^2 = F_0^2 ~, & \qquad & F_V^2 M_V^2 = F_A^2 M_A^2 ~.
\end{eqnarray} 
For the actual numerical values, we have used  $F_0=F_\pi$,
$c_d=c_m$,  $F_V=2 G_V$, $M_V=0.77$ GeV  and $M_S=1.4$ GeV.
The qualitative features of the numerical values of the $L_i^r$ are
well reproduced with resonance saturation, as shown in Table
\ref{tab:p4res}.  Two comments are in order.

\begin{itemize} 
\item The absolute values of $L_1, L_2, L_3$ for the $p^6$ fit are a
  bit smaller than the 
  vector meson contributions. As already noted in
  Refs.~\cite{Ecker:1988te,Ecker:1989yg}, the agreement improves if
  instead of using the KSFR relation $F_V=2 G_V$, the decay widths
  $\Gamma(\rho^0 \to e^+ e^-)$ and $\Gamma(\rho \to \pi\pi)$ are
    invoked to extract the coupling 
  constants $F_V$ and $G_V$. This type of fine-tuning is not our concern here.
\item The much-debated scalar resonance dominance works very well for
  $L_5$ and $L_8$. Essential for this agreement is the notion of the
  lightest scalar nonet that survives in the large-$N_c$ limit. We
  refer to Ref.~\cite{Cirigliano:2003yq} for a discussion in favour of
  an average scalar resonance mass in the vicinity of $M_S \simeq 1.4$
  GeV.
\end{itemize} 

\begin{table}[tb!]
\caption{\label{tab:p4res} Comparison of the fitted $L_i^r$ at NLO
  (Table \ref{tab:p4fits}) and NNLO (Table \ref{tab:p6fits}) with
  resonance saturation \cite{Ecker:1988te,Ecker:1989yg}. The numerical 
  estimates of the resonance contributions are based on
  Eqs.~(\ref{eq:reso}), with $F_0=F_\pi$, $c_d=c_m$, $F_V=2 G_V$,
  $M_V=0.77$ GeV  and $M_S=1.4$ GeV. All numerical values 
  are in units of $10^{-3}$.}
\begin{center}
\begin{tabular}{@{}crrcc@{}}
\hline\hline
& & & & \\[-.3cm] 
 & $O(p^4)$  &  $O(p^6)$ & R exchange & num. estimate  \\[.1cm]  
\hline \\[-.3cm] 
$L_1^r$ &   1.0(1) &   0.53(6) &  $\frac{G_V^2}{8 M_V^2}$ & 0.9
\\[.1cm] 
$L_2^r$ &   1.6(2) &   0.81(4) &  $\frac{G_V^2}{4 M_V^2}$ & 1.8
\\[.1cm] 
$L_3^r$ &   $-$ 3.8(3) & $-$ 3.07(20) &  $- \frac{3 G_V^2}{4 M_V^2}
+ \frac{c_d^2}{2 M_S^2}$ & $-$ 4.8 
\\[.1cm] 
$L_4^r$ &   0.0(3) &   0.3 &  0 & 0
\\[.1cm] 
$L_5^r$ &   1.2(1) &   1.01(6) & $ \frac{c_d c_m}{M_S^2}$  & 1.1
\\[.1cm] 
$L_6^r$ &   0.0(4) &   0.14(5) & 0  & 0
\\[.1cm] 
$L_7^r$ &   $-$ 0.3(2) &  $-$ 0.34(9) &  $- \frac{F_0^2}{48
  M_{\eta^\prime}^2}$ & $-$ 0.2
\\[.1cm] 
$L_8^r$ &  0.5(2) &   0.47(10) & $\frac{c_m^2}{2 M_S^2}$  & 0.54
\\[.1cm] 
$L_9^r$ & 6.9(7)  &  5.9(4) &  $\frac{F_V G_V}{2 M_V^2}$ & 7.2
\\[.1cm] 
$L_{10}^r$ & $-$5.2(1)  & $-$3.8(4) &  $- \frac{F_V^2}{4 M_V^2} + \frac{F_A^2}{4
  M_A^2}$ & $-$ 5.4
\\[.1cm] 
\hline
\end{tabular}
\end{center}
\end{table}

We now turn to the issue of resonance saturation of the NNLO LECs
$C_i^r$. The most general chiral resonance Lagrangian that can
generate chiral LECs up to $O(p^6)$ was constructed in
Ref.~\cite{Cirigliano:2006hb}. The corresponding chiral resonance
theory generates Green functions that interpolate between QCD and
\chpt. It is therefore natural to expect that resonance saturation
works qualitatively also for the $C_i^r$.

However, the situation is more complicated than at $O(p^4)$. First,
as discussed in the previous subsection, our knowledge of the
numerical values of the $C_i^r$ is still limited. Second, many more
couplings arise at $O(p^6)$, many of them related to double-resonance
exchange, which are essentially unknown. In the following, we will
therefore investigate relations for the $C_i^r$ that fulfill two 
conditions: These LECs contribute to observables used in our fits of
the $L_i^r$ and, secondly, the relations either involve only resonance
parameters that occurred already at $O(p^4)$ or they do not depend on
any parameters at all. A list of such relations was given in
Ref.~\cite{Cirigliano:2006hb} but not analysed there. Note that many of
the studies in this area assume that short-distance constraints should be
satisfied. This is not always possible with a finite number of resonances
\cite{Bijnens:2003rc} and in that case a choice of what is implemented
is necessary.

\begin{table}[tb!]
\caption{\label{tab:p6res} Relations among LECs $C_i^r$ from resonance
  exchange \cite{Cirigliano:2006hb,Kaiser:2007zz} that are either
  parameter-free or depend only on parameters 
  occurring already at $O(p^4)$. The values in the last column are
  taken from Table \ref{tab:Cafit} with the usual
  renormalization scale $\mu=0.77$ GeV. The numerical estimates are
  given in units of $10^{-6}$, with the same input values as in 
  Table~\ref{tab:p4res}. }
\begin{center}
\begin{tabular}{@{}lccc@{}}
\hline\hline
& & & \\[-.3cm] 
 LECs &  R exchange & num. estimate & fit value \\[.1cm]  
\hline \\[-.3cm] 
$C_1 + C_3 - C_4$ &  $- \frac{c_d^2 F_\pi^2}{4 M_S^4}$ & $-$ 1.2 & 1
\\[.1cm] 
$3 C_3 + C_4$ &  ${ \frac{G_V^2 F_\pi^2}{8 M_V^4}}$ &  13 & 27
\\[.1cm] 
$C_{12}$ &  $- \frac{c_d c_m F_\pi^2}{2 M_S^4}$ & $-$ 2.4 & $-$ 2.8
\\[.1cm] 
$C_{18}$ &  $- \frac{F_\pi^4}{48 M_{\eta^\prime}^4}$ & $-$ 1.8 & 0.6
\\[.1cm] 
$C_{19}$ &  $- \frac{F_\pi^4}{144 M_{\eta^\prime}^4}$ & $-$ 0.6 & $-$ 4.0
\\[.1cm] 
$C_{20}$ &  $\frac{F_\pi^4}{96 M_{\eta^\prime}^4}$ & 0.9 & 1.0
\\[.1cm] 
$C_{20} + 3 C_{21}$ &  0 & 0 & $-$ 0.4
\\[.1cm] 
$C_{32} + 3 C_{21}$ &  0 & 0 & 0.3
\\[.1cm] 
$C_{28} - C_{30}/2$ &  0 & 0 & $-$ 3.5
\\[.1cm] 
$C_1/12 - C_{28} + \frac{c_d}{c_m} C_{32}$ &  $- \frac{7 c_d^2
  F_\pi^2}{144 M_S^4} - \frac{G_V^2 F_\pi^2}{288 M_V^4} +
\frac{c_d}{c_m} \frac{F_\pi^4}{96 M_{\eta^\prime}^4}$ &  0.3 & 4.7  
\\[.1cm] 
\hline
\end{tabular}
\end{center}
\end{table}

In Table \ref{tab:p6res} we display the relations involving only
those $C_i^r$ that contribute to our observables. We make several
comments. 

\begin{itemize} 
\item The numerical estimates should be viewed with the naive
  dimensional estimate of NNLO LECs in mind ($C_0 = 40 \cdot
  10^{-6}$).
\item While the fit values in the last column of Table \ref{tab:p6res}
  refer to the usual renormalization scale ($\mu=0.77$ GeV), the
  numerical estimates from resonance exchange do not carry a scale
  dependence (leading order in $1/N_c$). For instance, for a scale
  $\mu=0.85$ GeV, the fit value for the combination $C_1 + C_3 - C_4$
  moves from 1 to $-$ 1.4, practically coinciding with the estimate
  from resonance exchange.
\item $C_1 + C_3 - C_4$ and $C_{12}$ are only sensitive to scalar
  exchange. As for $L_5$ and $L_8$ (see Table \ref{tab:p4res}), the
  predictions work surprisingly well. There is certainly no evidence
  for a failure of scalar resonance saturation.
\item The parameter-free relations involving $C_{20}$, $C_{21}$ and
  $C_{32}$ are not only well satisfied, but the involved LECs also
  seem to be dominated by $\eta^\prime$ exchange in accordance with
  large $N_c$ \cite{Kaiser:2007zz}.
\item Finally, we cannot compare our results directly with one of the
  most solid predictions for NNLO LECs, i.e. for $C_{88} - C_{90}$ in
  Eq.~(\ref{eq:8890}). However, as the last entry in Table \ref{tab:Cafit}
  documents, our values for the $C_i$ certainly do not contradict that
  prediction.
\end{itemize}   

\subsection{Convergence of \chpt\ for chiral $SU(3)$}
\label{sec:convergence}

We now study the convergence of a number of quantities for the fit BE14
and the free fit. This can be compared with the same study done in
Ref.~\cite{Bijnens:2011tb}.

The quantities are always in the order LO+NLO+NNLO and the main number is fit
BE14 while the number in brackets is from the free fit.
\begin{eqnarray}
{F_K/F_\pi} &=& {1 + 0.176(0.121) + 0.023(0.077)}\,,
\nonumber\\
F_\pi/F_0 &=& {1+0.208(0.313) + 0.088(0.127)}\,,  
\nonumber\\
M_\pi^2/M_{\pi\mathrm{phys}}^2 &=& {1.055(0.937) - 0.005(+ 0.107) - 0.050(-0.044)}\,,
\nonumber\\
M_K^2/M_{K\mathrm{phys}}^2 &=& {1.112(0.994) - 0.069(+ 0.022) - 0.043(-0.016)}\,,
\nonumber\\
M_\eta^2/M_{\eta\mathrm{phys}}^2 &=& {1.197(0.938) - 0.214(-0.076)+ 0.017(0.014)}\,.
\end{eqnarray}
The LO contribution to the masses is calculated from our NLO and NNLO 
results. The total higher-order corrections are very reasonable for all
the ratios listed above even though the NLO corrections are 
small in some cases.

The $\pi\pi$ scattering lengths show a very good convergence
for both:
\begin{eqnarray}
a^0_0 &=& 0.160+0.044(0.046)+0.012(0.012)\,,
\nonumber\\ 
a^2_0 &=& -0.0456+0.0016(0.0017)-0.0001(-0.0003)\,.
\end{eqnarray}
The $\pi K$ scattering lengths have a worse convergence:
\begin{eqnarray}
a^{1/2}_0 &=& 0.142+0.031(0.027)+0.051(0.057)\,,
\nonumber\\
a^{3/2}_0 &=& -0.071+0.007(0.005)+0.016(0.019)\,.
\end{eqnarray}

Finally, we present the convergence for the $K_{l4}$ form factors at
threshold: 
\begin{eqnarray}
f_s &=& 3.786 + 1.202(1.231) +0.717(0.688)\,,
\nonumber\\
g_p &=& 3.786 + 0.952(0.857) +0.212(0.309)\,.
\end{eqnarray}
Note that we have fitted the observables to within their errors, so all
higher-order contributions are included in the NNLO parts.
The overall picture is that the convergence is in line with
expectations for $n_f=3$ ChPT.

\subsection{Dynamical photons and leptons}
\label{sec:photons}

As shown in Ref.~\cite{Moussallam:1997xx}, the
electromagnetic LECs $K_i$ of the Lagrangian (\ref{eq:Le2p2}) obey
integral sum rule representations, generalizing the DGMLY sum rule
\cite{Das:1967it} for the $\pi^+$-$\pi^0$ mass difference. The integral
representations have the form of convolutions of pure QCD $n$-point
functions ($n \le 4$) with the free photon propagator. The
representations serve several purposes \cite{Moussallam:1997xx}: They
can be used to study the dependence of the $K_i^r$ on the chiral
renormalization scale, their gauge dependence and possible
short-distance ambiguities. The representations also lead to
model-independent relations among the LECs. Last but not least, they
allow for approximate determinations of the LECs by saturating the
integrals  in terms of resonance exchanges.
This is especially important in the present case because it is nearly
impossible to  determine the LECs $K_i$ from phenomenology.

In Ref.~\cite{Moussallam:1997xx} the method was applied to
$K_7$, \dots, $K_{13}$ involving two- and three-point functions only. 
$K_{14}$ multiplies a pure source term and is therefore
irrelevant for phenomenology. It
turns out that $K_7$ and $K_8$ are large-$N_c$ suppressed and are
therefore set to zero at the scale $\mu=M_\rho$. The remaining LECs
$K_9$, \dots, $K_{13}$ are all gauge dependent. In fact, 
$K_{9}$, \dots, $K_{12}$ also depend on the QCD renormalization scale
$\mu_{\rm SD}$ \cite{Moussallam:1997xx,Bijnens:1996kk}.
These LECs can therefore not be expressed separately in terms of
physical quantities but will occur only in certain combinations in
observables. For instance, the combination $K_{10}+K_{11}$ enters the
corrections of  $O(e^2 m_s)$ to Dashen's theorem \cite{Dashen:1969eg}.
Consequently, this combination is independent of the gauge parameter
$\xi$ and $\mu_{\rm SD}$, depending only on the chiral renormalization
scale $\mu$. 
With this proviso in mind, numerical values will be  
given in Feynman gauge ($\xi = 1$), for $\mu_{\rm SD}=1$ GeV and for
the usual chiral renormalization scale $\mu=M_\rho$. 
Note that in contrast to the strong LECs, the
chiral scale dependence already appears at leading order in
$1/N_c$. Other estimates exist in the calculations of the corrections
to Dashen's theorem \cite{Donoghue:1993hj,Bijnens:1993ae}. These use other
methods to estimate the intermediate- and short-distance
momentum regimes. In particular, Ref.~\cite{Bijnens:1996kk} 
treated the intermediate-distance dynamics with the ENJL model and the
short-distance part with perturbative QCD and factorization.
Using the latter method, it is also very clear why the LECs are gauge
and QCD scale dependent \cite{Bijnens:1996kk,Gasser:2003hk}.

The more complicated case of four-point functions in the sum rule
representations of $K_1$, \dots, $K_6$ was investigated in
Ref.~\cite{Ananthanarayan:2004qk}. It turns out that all these LECs
are gauge independent\footnote{The $\beta$-functions of the $K_i
~(i=1,\dots,14)$ in a general covariant gauge were calculated in
 Ref.~\cite{Agadjanov:2013lra}.}. Independently of the single-resonance
approximation, one can derive the following large-$N_c$ relations
\cite{Bijnens:1996kk,Ananthanarayan:2004qk}: 
\begin{eqnarray}
\label{eq:Kirel}
K_3^r = - K_1^r~, \qquad & \qquad K_4^r = 2 K_2^r~.
\end{eqnarray} 
In Table \ref{tab:Kinum} we collect numerical results for the
$K_i^r(M_\rho)$ on the basis of the sum rule representations of
Refs.~\cite{Moussallam:1997xx,Ananthanarayan:2004qk}. As the authors
emphasize, uncertainties of the numerical predictions are difficult to
estimate quantitatively even for physically relevant combinations:
Both the large-$N_c$ approximation and the 
single-resonance assumption (except for the relations
(\ref{eq:Kirel})) should be kept in mind. The values in the other
approaches have
the same order of magnitude but differ in the detailed predictions.

\begin{table}[tb!]
\caption{\label{tab:Kinum} 
Numerical values of the LECs $K_i^r(M_\rho)$ in  units
of $10^{-3}$ \cite{Moussallam:1997xx,Ananthanarayan:2004qk}. The
gauge-dependent LECs (see text) are given in Feynman 
gauge ($\xi=1$). The QCD renormalization scale is set to $\mu_{\rm
  SD}=1$ GeV.} 
\begin{center}
\begin{tabular}{@{}ccccc@{}}
\hline\hline 
& & & & \\[-.3cm] 
 LEC  &   & \hspace*{.5cm}  & LEC   &  
\\[.1cm] 
\hline
& & & & \\[-.3cm]
 $K_{1}^r$ & $-2.7$ & &
  $K_{7}^r$ & $\simeq 0$ 
   \\
 $K_{2}^r$ & $0.7$ & &
$K_{8}^r$  & $\simeq 0$
  \\
 $K_{3}^r$ & $2.7$ & &
$K_{10}^r$  & $7.5$ 
   \\
 $K_{4}^r$ & $1.4$ & &
$K_{11}^r$  & $1.3$   
   \\
 $K_{5}^r$ & $11.6$ & &
$K_{12}^r$  & $-4.2$ 
   \\
 $K_{6}^r$ & $2.8$ & &
$K_{13}^r$  & $4.7$ 
   \\
\hline
\end{tabular}
\end{center}
\end{table}

Finally, we turn to the case of dynamical photons and leptons for the
calculation of
radiative corrections in semileptonic meson decays. The
corresponding LECs $X_1$, \dots, $X_7$ are defined in the Lagrangian
(\ref{eq:Llept}). Actually, neither $X_4$ nor $X_7$ are
phenomenologically relevant. In analogy to the formalism set up
for the $K_i$, Descotes-Genon and Moussallam
\cite{DescotesGenon:2005pw} established integral representations for
all $X_i$ with the help of a two-step matching procedure (Standard
Model ~$\rightarrow$~  Fermi theory   ~$\rightarrow$~ \chpt). 
These representations furnish numerical estimates of the LECs, once
the chiral Green functions are approximated with the help of
large-$N_c$-motivated models. 

Again as for the $K_i$, the integral representations also allow the
derivation of non-trivial relations among the $X_i$. Independently of
any model for the two- and three-point functions involved, the
following relations hold \cite{DescotesGenon:2005pw}:
\begin{eqnarray}
\label{eq:Xirel}
X_2^r = \left(X_3^r + \frac{3}{32 \pi^2}\right)/4, \qquad & \qquad X_5^r = - 2
X_2^r 
\end{eqnarray}      
so that in fact only three independent LECs remain to be estimated
with specific models.

In Table \ref{tab:Xinum} we collect the numerical estimates for the
LECs $X_1$, $X_2$, $X_3$ and $X_5$, putting $X_6$ aside for the
moment. Except for the model-independent relations (\ref{eq:Xirel}),
the estimates are based on a minimal resonance model with a single
multiplet of vector and axial-vector resonances each (only $V$ and $A$
spectral functions are involved).

\begin{table}[tb!] 
\caption{\label{tab:Xinum} 
Numerical values of the LECs $X_i^r(M_\rho)$
($i=1,2,3,5$) for $M_A = \sqrt{2} M_\rho$ 
\cite{DescotesGenon:2005pw}.} 
\begin{center}
\begin{tabular}{@{}cccc@{}}
\hline\hline 
& & &  \\[-.3cm] 
  $10^{3} X_1^r$ &  $10^{3} X_2^r$ &  $10^{3} X_3^r$
 &   $10^{3} X_5^r$   
 \\
\hline
& & &  \\[-.3cm] 
 $- 3.7$ & $3.6$ & $5.0$ & $- 7.2$
   \\[4pt]
\hline  
\end{tabular}
\end{center}
\end{table}

The LEC $X_6$ plays a special role because it cannot be determined
from the matching conditions in the same way as $X_1$, $X_2$, $X_3$
and $X_5$. By looking at explicit
calculations of radiative corrections of semileptonic decays, one
verifies that $X_6$ and $K_{12}$ always appear in the same combination 
$X_6^{\rm phys}:= X_6 - 4 K_{12}$ related to wave-function
renormalization. 

It is convenient to write $X_6^{\rm phys}$ as the sum of two contributions:
\begin{equation} 
\label{eq:X6split}
X_6^{\rm phys}(\mu) = X_{6,{\rm SD}}^{\rm phys} + 
                   {\tilde X}_6^{\rm phys}(\mu)~,
\end{equation} 
where the short-distance contribution is given by \cite{Sirlin:1981ie}
\begin{equation} 
\label{eq:X6SD}
 X_{6,{\rm SD}}^{\rm phys} = - \frac{1}{2 \pi^2} \log{\frac{M_Z}{M_V}}
\end{equation}
and the remainder has the following form in the single-resonance
approximation \cite{DescotesGenon:2005pw}:
\begin{equation} 
\label{eq:X6rest}
{\tilde X}_6^{\rm phys}(\mu) = \frac{1}{16 \pi^2} \left(3
\log{\frac{\mu^2}{M_V^2}} + \frac{1}{2} 
\log{\frac{M_A^2}{M_V^2}} - \displaystyle\frac{3 M_V^2 + M_A^2}{(4 \pi
  F_0)^2} + \frac{7}{2} \right)~.
\end{equation} 
Summing up powers of electroweak logarithms and adding a small correction of
$O(\alpha_s)$ \cite{Marciano:1993sh},
\begin{eqnarray} 
\label{eq:resum}
 X_{6,{\rm SD}}^{\rm phys} = - 0.2419 \qquad   & \longrightarrow & \qquad
 {\bar X}_{6,{\rm SD}}^{\rm phys} = - 0.2527~.
\end{eqnarray} 
By convention, the short-distance contribution is factorized and
appears in a universal multiplicative factor
\begin{equation}
\label{eq:SEW}
S_{\rm EW} = 1- e^2 {\bar X}_{6,{\rm SD}}^{\rm phys} = 1.0232
\end{equation} 
in all radiatively corrected semileptonic decay rates. The subdominant
remainder ${\tilde X}_6^{\rm phys}(\mu)$, on the other hand, combines
with other terms in the decay amplitude to yield a scale-independent
expression. In the model of Ref.~\cite{DescotesGenon:2005pw} one
obtains (with $M_A = \sqrt{2} M_\rho$)
\begin{equation} 
{\tilde X}_6^{\rm phys}(M_\rho) = 0.0104~.
\end{equation}
  
\section{CONCLUSIONS AND FINAL RESULTS}

\chpt\ as a nonrenormalizable effective field theory requires reliable
information on many of its coupling constants in order to arrive at
meaningful predictions. 
In this review we have collected the available knowledge of the
low-energy constants in mesonic \chpt, emphasizing the
chiral Lagrangians for the strong interactions. 

For \nftwo\ \chpt\ the NLO LECs are by
now quite well known from phenomenology with the exception of $\bar
l_3$, for which one should turn to the lattice.
The values for the $\bar l_i$ are summarized in
Eqs. (\ref{eq:l1l2},\ref{eq:l4},\ref{eq:l5},\ref{eq:l6b},\ref{eq:l3l4}).
The convergence for most of the quantities studied is excellent.
The phenomenological knowledge of the NNLO LECs given in (\ref{eq:ci})
is still rather modest. Here we have restricted ourselves to
quoting published results.

For \nfthree\ \chpt\ we have given a short overview of the different types
of chiral Lagrangians. We then concentrated on a new fit of the
NLO LECs in the strong sector, using all available information about the
NNLO LECs. 
With reasonable values for the $C_i^r$ a good fit can be obtained with
satisfactory convergence for many physical quantities. However, one
should keep in mind that determining $L_4^r$ from 
continuum phenomenology is very difficult. Lattice results and
large-$N_c$ arguments suggest $|L_4^r(M_\rho)| \le 3 \cdot 10^{-3}$. This
leads to our main new NNLO fit for the $L_i^r$ given in column BE14 in
Table \ref{tab:p6fits}. We have emphasized that fit BE14 should always
be considered together with the associated set of $C_i^r$ values in
Table \ref{tab:Cafit}. Although the changes in the $L_i^r$ are
non-negligible when going from NLO to NNLO, the pattern is
quite stable. Another interesting feature is that requiring a
small  $|L_4^r(M_\rho)|$ leads automatically to small values of
$|2 L_1^r(M_\rho) -  L_2^r(M_\rho)|$ and $|L_6^r(M_\rho)|$ in
accordance with large $N_c$. We have compared our findings with
available results from lattice studies.  

Quite generally, LECs parametrize the physics at shorter
distances. We have taken a fresh look at the evidence for resonance
saturation of the strong LECs, confirming the qualitative agreement
for the $L_i^r$ and finding new evidence also for some of the
$C_i^r$. In the few cases where only scalar resonances contribute,
resonance saturation seems to work as well at least qualitatively for
both NLO and NNLO LECs. In addition, our preferred values of the
$C_i^r$ are consistent with $\eta^\prime$ exchange in accordance with
large $N_c$. 

Although not as impressive as for \nftwo, most of the observables used
for our fits show a reasonable ``convergence'' also for \nfthree, once
this pattern is enforced for the meson masses.

Finally, we have reviewed the status of the LECs occurring in the
chiral Lagrangians with dynamical photons and leptons relevant for
radiative corrections. Although phenomenology is not of much help for
a determination of those LECs, different theoretical approaches have
led to a consistent picture for all NLO LECs. The interplay between
intermediate- and short-distance physics is well under control. 

\section*{Acknowledgements}

We thank H. Neufeld, I. Jemos and A. Pich for helpful discussions.
This work is supported in part by the European Community-Research
Infrastructure Integrating Activity ``Study of Strongly Interacting Matter''
(HadronPhysics3, Grant Agreement No. 283286)
and the Swedish Research Council grants 621-2011-5080 and 621-2013-4287.


\begin{thebibliography}{10}

\bibitem{Weinberg:1978kz}
  S.~Weinberg,
  Physica A {\bf 96} (1979) 327.



\bibitem{Gasser:1983yg}
  J.~Gasser and H.~Leutwyler,
  Annals Phys.\  {\bf 158} (1984) 142.



\bibitem{Gasser:1984gg}
  J.~Gasser and H.~Leutwyler,
  Nucl.\ Phys.\ B {\bf 250} (1985) 465.



\bibitem{Riggenbach:1990zp}
  C.~Riggenbach, J.~Gasser, J.~F.~Donoghue and B.~R.~Holstein,
  Phys.\ Rev.\ D {\bf 43} (1991) 127.



\bibitem{Bijnens:1989mr}
  J.~Bijnens,
  Nucl.\ Phys.\ B {\bf 337} (1990) 635.



\bibitem{Bijnens:1994ie}
  J.~Bijnens, G.~Colangelo and J.~Gasser,
  Nucl.\ Phys.\ B {\bf 427} (1994) 427
  [hep-ph/9403390].



\bibitem{Amoros:2000mc}
  G.~Amor{\'o}s, J.~Bijnens and P.~Talavera,
  Nucl.\ Phys.\ B {\bf 585} (2000) 293
   [Erratum ibid.\ B {\bf 598} (2001) 665]
  [hep-ph/0003258].



\bibitem{Amoros:2001cp}
  G.~Amor{\'o}s, J.~Bijnens and P.~Talavera,
  Nucl.\ Phys.\ B {\bf 602} (2001) 87
  [hep-ph/0101127].



\bibitem{Kaplan:1986ru}
  D.~B.~Kaplan and A.~V.~Manohar,
  Phys.\ Rev.\ Lett.\  {\bf 56} (1986) 2004.



\bibitem{Bijnens:2011tb}
  J.~Bijnens and I.~Jemos,
  Nucl.\ Phys.\ B {\bf 854} (2012) 631
  [arXiv:1103.5945 [hep-ph]].



\bibitem{Cirigliano:2011ny}
  V.~Cirigliano {\it et al.}, 
  Rev.\ Mod.\ Phys.\  {\bf 84} (2012) 399
  [arXiv:1107.6001 [hep-ph]].



\bibitem{Colangelo:2010et}
  G.~Colangelo  {\it et al.}, 
  Eur.\ Phys.\ J.\ C {\bf 71} (2011) 1695
  [arXiv:1011.4408 [hep-lat]].



\bibitem{Aoki:2013ldr}
  S.~Aoki {\it et al.}, 
  arXiv:1310.8555 [hep-lat].



\bibitem{Leutwyler:1993iq}
  H.~Leutwyler,
  Annals Phys.\  {\bf 235} (1994) 165
  [hep-ph/9311274].



\bibitem{Scherer:2012xha}
  S.~Scherer and M.~R.~Schindler,
  Lect.\ Notes Phys.\  {\bf 830} (2012) 1.



\bibitem{Ecker:1994gg}
  G.~Ecker,
  Prog.\ Part.\ Nucl.\ Phys.\  {\bf 35} (1995) 1
  [hep-ph/9501357].



\bibitem{Pich:1995bw}
  A.~Pich,
  Rept.\ Prog.\ Phys.\  {\bf 58} (1995) 563
  [hep-ph/9502366].



\bibitem{Bijnens:2006zp}
  J.~Bijnens,
  Prog.\ Part.\ Nucl.\ Phys.\  {\bf 58} (2007) 521
  [hep-ph/0604043].



\bibitem{Bijnens:1999sh}
  J.~Bijnens, G.~Colangelo and G.~Ecker,
  JHEP {\bf 9902} (1999) 020
  [hep-ph/9902437].



\bibitem{Fearing:1994ga}
  H.~W.~Fearing and S.~Scherer,
  Phys.\ Rev.\ D {\bf 53} (1996) 315
  [hep-ph/9408346].



\bibitem{Haefeli:2007ty}
  C.~Haefeli, M.~A.~Ivanov, M.~Schmid and G.~Ecker,
  arXiv:0705.0576 [hep-ph].



\bibitem{Bijnens:1997vq}
  J.~Bijnens {\it et al.}, 
  Nucl.\ Phys.\ B {\bf 508} (1997) 263
   [Erratum ibid.\ B {\bf 517} (1998) 639]
  [hep-ph/9707291].



\bibitem{Bijnens:1999hw}
  J.~Bijnens, G.~Colangelo and G.~Ecker,
  Annals Phys.\  {\bf 280} (2000) 100
  [hep-ph/9907333].



\bibitem{Ecker:1988te}
  G.~Ecker, J.~Gasser, A.~Pich and E.~de Rafael,
  Nucl.\ Phys.\ B {\bf 321} (1989) 311.



\bibitem{Moussallam:1998za}
  B.~Moussallam,
  Eur.\ Phys.\ J.\ C {\bf 6} (1999) 681
  [hep-ph/9804271].



\bibitem{Urech:1994hd}
  R.~Urech,
  Nucl.\ Phys.\ B {\bf 433} (1995) 234
  [hep-ph/9405341].



\bibitem{Knecht:1999ag}
  M.~Knecht, H.~Neufeld, H.~Rupertsberger and P.~Talavera,
  Eur.\ Phys.\ J.\ C {\bf 12} (2000) 469
  [hep-ph/9909284].



\bibitem{Wess:1971yu}
  J.~Wess and B.~Zumino,
  Phys.\ Lett.\ B {\bf 37} (1971) 95.



\bibitem{Witten:1983tw}
  E.~Witten,
  Nucl.\ Phys.\ B {\bf 223} (1983) 422.



\bibitem{Ebertshauser:2001nj}
  T.~Ebertshauser, H.~W.~Fearing and S.~Scherer,
  Phys.\ Rev.\ D {\bf 65} (2002) 054033
  [hep-ph/0110261].



\bibitem{Bijnens:2001bb}
  J.~Bijnens, L.~Girlanda and P.~Talavera,
  Eur.\ Phys.\ J.\ C {\bf 23} (2002) 539
  [hep-ph/0110400].



\bibitem{Moussallam:1997xx}
  B.~Moussallam,
  Nucl.\ Phys.\ B {\bf 504} (1997) 381
  [hep-ph/9701400].



\bibitem{Knecht:2001xc}
  M.~Knecht and A.~Nyffeler,
  Eur.\ Phys.\ J.\ C {\bf 21} (2001) 659
  [hep-ph/0106034].



\bibitem{RuizFemenia:2003hm}
  P.~D.~Ruiz-Femen{\'i}a, A.~Pich and J.~Portol{\'e}s,
  JHEP {\bf 0307} (2003) 003
  [hep-ph/0306157].



\bibitem{Ananthanarayan:2002kj}
  B.~Ananthanarayan and B.~Moussallam,
  JHEP {\bf 0205} (2002) 052
  [hep-ph/0205232].



\bibitem{Cronin:1967jq}
  J.~A.~Cronin,
  Phys.\ Rev.\  {\bf 161} (1967) 1483.



\bibitem{Kambor:1989tz}
  J.~Kambor, J.~H.~Missimer and D.~Wyler,
  Nucl.\ Phys.\ B {\bf 346} (1990) 17.



\bibitem{Ecker:1992de}
  G.~Ecker, J.~Kambor and D.~Wyler,
  Nucl.\ Phys.\ B {\bf 394} (1993) 101.



\bibitem{Bijnens:2004ai}
  J.~Bijnens and F.~Borg,
  Eur.\ Phys.\ J.\ C {\bf 40} (2005) 383
  [hep-ph/0501163].



\bibitem{D'Ambrosio:1997tb}
  G.~D'Ambrosio and J.~Portol{\'e}s,
  Nucl.\ Phys.\ B {\bf 533} (1998) 494
  [hep-ph/9711211].



\bibitem{Pallante:2001he}
  E.~Pallante, A.~Pich and I.~Scimemi,
  Nucl.\ Phys.\ B {\bf 617} (2001) 441
  [hep-ph/0105011].



\bibitem{Cirigliano:2003gt}
  V.~Cirigliano, G.~Ecker, H.~Neufeld and A.~Pich,
  Eur.\ Phys.\ J.\ C {\bf 33} (2004) 369
  [hep-ph/0310351].



\bibitem{Bardeen:1986uz}
  W.~A.~Bardeen, A.~J.~Buras and J.-M.~G{\'e}rard,
  Nucl.\ Phys.\ B {\bf 293} (1987) 787.



\bibitem{Bijnens:1998ee}
  J.~Bijnens and J.~Prades,
  JHEP {\bf 9901} (1999) 023
  [hep-ph/9811472].



\bibitem{Buras:2014maa}
  A.~J.~Buras, J.-M.~G{\'e}rard and W.~A.~Bardeen,
  arXiv:1401.1385 [hep-ph].



\bibitem{Bijnens:1999zn}
  J.~Bijnens and J.~Prades,
  JHEP {\bf 0001} (2000) 002
  [hep-ph/9911392].



\bibitem{Peris:1998nj}
  S.~Peris, M.~Perrottet and E.~de Rafael,
  JHEP {\bf 9805} (1998) 011
  [hep-ph/9805442].



\bibitem{Bijnens:1983ye}
  J.~Bijnens and M.~B.~Wise,
  Phys.\ Lett.\ B {\bf 137} (1984) 245.



\bibitem{Grinstein:1985ut}
  B.~Grinstein, S.-J.~Rey and M.~B.~Wise,
  Phys.\ Rev.\ D {\bf 33} (1986) 1495.



\bibitem{Cirigliano:1999hj}
  V.~Cirigliano, J.~F.~Donoghue and E.~Golowich,
  Phys.\ Rev.\ D {\bf 61} (2000) 093002
  [hep-ph/9909473].



\bibitem{Bijnens:2001ps}
  J.~Bijnens, E.~G{\'a}miz and J.~Prades,
  JHEP {\bf 0110} (2001) 009
  [hep-ph/0108240].



\bibitem{Ecker:2000zr}
  G.~Ecker {\it et al.}, 
  Nucl.\ Phys.\ B {\bf 591} (2000) 419
  [hep-ph/0006172].



\bibitem{Burgi:1996qi}
  U.~B{\"u}rgi,
  Nucl.\ Phys.\ B {\bf 479} (1996) 392
  [hep-ph/9602429].



\bibitem{Bijnens:1995yn}
  J.~Bijnens {\it et al.}, 
  Phys.\ Lett.\ B {\bf 374} (1996) 210
  [hep-ph/9511397].



\bibitem{Bijnens:1998fm}
  J.~Bijnens, G.~Colangelo and P.~Talavera,
  JHEP {\bf 9805} (1998) 014
  [hep-ph/9805389].



\bibitem{Bijnens:1996wm}
  J.~Bijnens and P.~Talavera,
  Nucl.\ Phys.\ B {\bf 489} (1997) 387
  [hep-ph/9610269].



\bibitem{Beringer:1900zz}
  J.~Beringer {\it et al.}  [Particle Data Group Collaboration],
  Phys.\ Rev.\ D {\bf 86} (2012) 010001.



\bibitem{Roy:1971tc}
  S.~M.~Roy,
  Phys.\ Lett.\ B {\bf 36} (1971) 353.



\bibitem{Ananthanarayan:2000ht}
  B.~Ananthanarayan, G.~Colangelo, J.~Gasser and H.~Leutwyler,
  Phys.\ Rept.\  {\bf 353} (2001) 207
  [hep-ph/0005297].



\bibitem{Colangelo:2001df}
  G.~Colangelo, J.~Gasser and H.~Leutwyler,
  Nucl.\ Phys.\ B {\bf 603} (2001) 125
  [hep-ph/0103088].



\bibitem{DescotesGenon:2001tn}
  S.~Descotes-Genon, N.~H.~Fuchs, L.~Girlanda and J.~Stern,
  Eur.\ Phys.\ J.\ C {\bf 24} (2002) 469
  [hep-ph/0112088].



\bibitem{GarciaMartin:2011cn}
  R.~Garcia-Martin {\it et al.}, 
  Phys.\ Rev.\ D {\bf 83} (2011) 074004
  [arXiv:1102.2183 [hep-ph]].



\bibitem{Batley:2007zz}
  J.~R.~Batley {\it et al.}  [NA48/2 Collaboration],
  Eur.\ Phys.\ J.\ C {\bf 54} (2008) 411.



\bibitem{Batley:2005ax}
  J.~R.~Batley {\it et al.}  [NA48/2 Collaboration],
  Phys.\ Lett.\ B {\bf 633} (2006) 173
  [hep-ex/0511056].



\bibitem{Adeva:2005pg}
  B.~Adeva {\it et al.}  [DIRAC Collaboration],
  Phys.\ Lett.\ B {\bf 619} (2005) 50
  [hep-ex/0504044].



\bibitem{Donoghue:1990xh}
  J.~F.~Donoghue, J.~Gasser and H.~Leutwyler,
  Nucl.\ Phys.\ B {\bf 343} (1990) 341.



\bibitem{Moussallam:1999aq}
  B.~Moussallam,
  Eur.\ Phys.\ J.\ C {\bf 14} (2000) 111
  [hep-ph/9909292].



\bibitem{Ananthanarayan:2004xy}
  B.~Ananthanarayan {\it et al.}, 
  Phys.\ Lett.\ B {\bf 602} (2004) 218
  [hep-ph/0409222].



\bibitem{Amendolia:1986wj}
  S.~R.~Amendolia {\it et al.}  [NA7 Collaboration],
  Nucl.\ Phys.\ B {\bf 277} (1986) 168.



\bibitem{Ananthanarayan:2013dpa}
  B.~Ananthanarayan, I.~Caprini, D.~Das and I.~Sentitemsu Imsong,
  Eur.\ Phys.\ J.\ C {\bf 73} (2013) 2520
  [arXiv:1302.6373 [hep-ph]].



\bibitem{GonzalezAlonso:2008rf}
  M.~Gonz{\'a}lez-Alonso, A.~Pich and J.~Prades,
  Phys.\ Rev.\ D {\bf 78} (2008) 116012
  [arXiv:0810.0760 [hep-ph]].



\bibitem{Bychkov:2008ws}
  M.~Bychkov {\it et al.}, 
  Phys.\ Rev.\ Lett.\  {\bf 103} (2009) 051802
  [arXiv:0804.1815 [hep-ex]].



\bibitem{Bijnens:2009zd}
  J.~Bijnens and I.~Jemos,
  Eur.\ Phys.\ J.\ C {\bf 64} (2009) 273
  [arXiv:0906.3118 [hep-ph]].



\bibitem{Baron:2011sf}
  R.~Baron {\it et al.}  [ETM Collaboration],
  PoS LATTICE {\bf 2010} (2010) 123
  [arXiv:1101.0518 [hep-lat]].



\bibitem{Arthur:2012opa}
  R.~Arthur {\it et al.}  [RBC and UKQCD Collaborations],
  Phys.\ Rev.\ D {\bf 87} (2013) 094514
  [arXiv:1208.4412 [hep-lat]].



\bibitem{Bazavov:2010yq}
  A.~Bazavov {\it et al.}, 
  PoS LATTICE {\bf 2010} (2010) 083
  [arXiv:1011.1792 [hep-lat]].



\bibitem{Beane:2011zm}
  S.~R.~Beane {\it et al.}, 
  Phys.\ Rev.\ D {\bf 86} (2012) 094509
  [arXiv:1108.1380 [hep-lat]].



\bibitem{Borsanyi:2012zv}
  S.~Bors{\'a}nyi {\it et al.}, 
  Phys.\ Rev.\ D {\bf 88} (2013) 014513
  [arXiv:1205.0788 [hep-lat]].



\bibitem{Baron:2009wt}
  R.~Baron {\it et al.}  [ETM Collaboration],
  JHEP {\bf 1008} (2010) 097
  [arXiv:0911.5061 [hep-lat]].



\bibitem{Durr:2013goa}
  S.~D{\"u}rr {\it et al.}, 
  arXiv:1310.3626 [hep-lat].



\bibitem{DescotesGenon:1999uh}
  S.~Descotes-Genon, L.~Girlanda and J.~Stern,
  JHEP {\bf 0001} (2000) 041
  [hep-ph/9910537].


\bibitem{'tHooft:1973jz}
  G.~'t Hooft,
  Nucl.\ Phys.\ B {\bf 72} (1974) 461.


\bibitem{Peris:1994dh}
  S.~Peris and E.~de Rafael,
  Phys.\ Lett.\ B {\bf 348} (1995) 539
  [hep-ph/9412343].



\bibitem{Ecker:2013pba}
  G.~Ecker, P.~Masjuan and H.~Neufeld,
  Eur.\ Phys.\ J.\ C {\bf 74} (2014) 2748
  [arXiv:1310.8452 [hep-ph]].



\bibitem{Amoros:1999dp}
  G.~Amor{\'o}s, J.~Bijnens and P.~Talavera,
  Nucl.\ Phys.\ B {\bf 568} (2000) 319
  [hep-ph/9907264].



\bibitem{Gasser:1984ux}
  J.~Gasser and H.~Leutwyler,
  Nucl.\ Phys.\ B {\bf 250} (1985) 517.



\bibitem{Bijnens:2003xg}
  J.~Bijnens and P.~Dhonte,
  JHEP {\bf 0310} (2003) 061
  [hep-ph/0307044].



\bibitem{Bijnens:2004eu}
  J.~Bijnens, P.~Dhonte and P.~Talavera,
  JHEP {\bf 0401} (2004) 050
  [hep-ph/0401039].



\bibitem{Bernard:1990kw}
  V.~Bernard, N.~Kaiser and U.~G.~Mei{\ss}ner,
  Nucl.\ Phys.\ B {\bf 357} (1991) 129.



\bibitem{Bijnens:2004bu}
  J.~Bijnens, P.~Dhonte and P.~Talavera,
  JHEP {\bf 0405} (2004) 036
  [hep-ph/0404150].



\bibitem{Buettiker:2003pp}
  P.~B{\"u}ttiker, S.~Descotes-Genon and B.~Moussallam,
  Eur.\ Phys.\ J.\ C {\bf 33} (2004) 409
  [hep-ph/0310283].



\bibitem{Batley:2012rf}
  J.~R.~Batley {\it et al.}  [NA48/2 Collaboration],
  Phys.\ Lett.\ B {\bf 715} (2012) 105
  [arXiv:1206.7065 [hep-ex]].



\bibitem{Bijnens:1996kk}
  J.~Bijnens and J.~Prades,
  Nucl.\ Phys.\ B {\bf 490} (1997) 239
  [hep-ph/9610360].



\bibitem{Bijnens:2002hp}
  J.~Bijnens and P.~Talavera,
  JHEP {\bf 0203} (2002) 046
  [hep-ph/0203049].



\bibitem{Gasser:2007sg}
  J.~Gasser, C.~Haefeli, M.~A.~Ivanov and M.~Schmid,
  Phys.\ Lett.\ B {\bf 652} (2007) 21
  [arXiv:0706.0955 [hep-ph]].



\bibitem{Jiang:2009uf}
  S.-Z.~Jiang, Y.~Zhang, C.~Li and Q.~Wang,
  Phys.\ Rev.\ D {\bf 81} (2010) 014001
  [arXiv:0907.5229 [hep-ph]].



\bibitem{Durr:1999dp}
  S.~D{\"u}rr and J.~Kambor,
  Phys.\ Rev.\ D {\bf 61} (2000) 114025
  [hep-ph/9907539].



\bibitem{Boito:2013qea}
  D.~Boito {\it et al.}, 
  arXiv:1311.6679 [hep-ph].



\bibitem{Bijnens:2003uy}
  J.~Bijnens and P.~Talavera,
  Nucl.\ Phys.\ B {\bf 669} (2003) 341
  [hep-ph/0303103].



\bibitem{Jamin:2004re}
  M.~Jamin, J.~A.~Oller and A.~Pich,
  JHEP {\bf 0402} (2004) 047
  [hep-ph/0401080].



\bibitem{Kampf:2006bn}
  K.~Kampf and B.~Moussallam,
  Eur.\ Phys.\ J.\ C {\bf 47} (2006) 723
  [hep-ph/0604125].



\bibitem{Prades:2007ud}
  J.~Prades,
  PoS KAON {\bf } (2008) 022
  [arXiv:0707.1789 [hep-ph]].



\bibitem{Unterdorfer:2008zz}
  R.~Unterdorfer and H.~Pichl,
  Eur.\ Phys.\ J.\ C {\bf 55} (2008) 273
  [arXiv:0801.2482 [hep-ph]].



\bibitem{Masjuan:2008fr}
  P.~Masjuan and S.~Peris,
  Phys.\ Lett.\ B {\bf 663} (2008) 61
  [arXiv:0801.3558 [hep-ph]].



\bibitem{Bernard:2009ds}
  V.~Bernard and E.~Passemar,
  JHEP {\bf 1004} (2010) 001
  [arXiv:0912.3792 [hep-ph]].



\bibitem{Boito:2012nt}
  D.~Boito {\it et al.}, 
  Phys.\ Rev.\ D {\bf 87} (2013) 094008
  [arXiv:1212.4471 [hep-ph]].



\bibitem{Cirigliano:2004ue}
  V.~Cirigliano {\it et al.}, 
  Phys.\ Lett.\ B {\bf 596} (2004) 96
  [hep-ph/0404004].



\bibitem{Cirigliano:2005xn}
  V.~Cirigliano {\it et al.}, 
  JHEP {\bf 0504} (2005) 006
  [hep-ph/0503108].



\bibitem{Rosell:2006dt}
  I.~Rosell, J.~J.~Sanz-Cillero and A.~Pich,
  JHEP {\bf 0701} (2007) 039
  [hep-ph/0610290].



\bibitem{Cirigliano:2006hb}
  V.~Cirigliano {\it et al.}, 
  Nucl.\ Phys.\ B {\bf 753} (2006) 139
  [hep-ph/0603205].



\bibitem{Kaiser:2007zz}
  R.~Kaiser,
  Nucl.\ Phys.\ Proc.\ Suppl.\  {\bf 174} (2007) 97.



\bibitem{Pich:2008jm}
  A.~Pich, I.~Rosell and J.~J.~Sanz-Cillero,
  JHEP {\bf 0807} (2008) 014
  [arXiv:0803.1567 [hep-ph]].



\bibitem{SanzCillero:2009ap}
  J.~J.~Sanz-Cillero and J.~Trnka,
  Phys.\ Rev.\ D {\bf 81} (2010) 056005
  [arXiv:0912.0495 [hep-ph]].



\bibitem{Pich:2010sm}
  A.~Pich, I.~Rosell and J.~J.~Sanz-Cillero,
  JHEP {\bf 1102} (2011) 109
  [arXiv:1011.5771 [hep-ph]].



\bibitem{Colangelo:2012ipa}
  P.~Colangelo, J.~J.~Sanz-Cillero and F.~Zuo,
  JHEP {\bf 1211} (2012) 012
  [arXiv:1207.5744 [hep-ph]].



\bibitem{Golterman:2014nua}
  M.~Golterman, K.~Maltman and S.~Peris,
  Phys.\ Rev.\ D {\bf 89} (2014) 054036
  [arXiv:1402.1043 [hep-ph]].



\bibitem{Bazavov:2010hj}
  A.~Bazavov {\it et al.}  [MILC Collaboration],
  PoS LATTICE {\bf 2010} (2010) 074
  [arXiv:1012.0868 [hep-lat]].



\bibitem{Ecker:2010nc}
  G.~Ecker, P.~Masjuan and H.~Neufeld,
  Phys.\ Lett.\ B {\bf 692} (2010) 184
  [arXiv:1004.3422 [hep-ph]].



\bibitem{Bazavov:2012cd}
  A.~Bazavov {\it et al.}, 
  Phys.\ Rev.\ D {\bf 87} (2013) 073012
  [arXiv:1212.4993 [hep-lat]].


\bibitem{Boyle:2014pja}
  P.~A.~Boyle {\it et al.},
  Phys.\ Rev.\ D {\bf 89} (2014) 094510
  [arXiv:1403.6729 [hep-ph]].


\bibitem{Bernard:2007tk}
  V.~Bernard and E.~Passemar,
  Phys.\ Lett.\ B {\bf 661} (2008) 95
  [arXiv:0711.3450 [hep-ph]].



\bibitem{Geng:2003mt}
  C.~Q.~Geng, I.-L.~Ho and T.~H.~Wu,
  Nucl.\ Phys.\ B {\bf 684} (2004) 281
  [hep-ph/0306165].



\bibitem{Moussallam:2000zf}
  B.~Moussallam,
  JHEP {\bf 0008} (2000) 005
  [hep-ph/0005245].



\bibitem{Bordes:2012ud}
  J.~Bordes {\it et al.}, 
  JHEP {\bf 1210} (2012) 102
  [arXiv:1208.1159 [hep-ph]].



\bibitem{Bazavov:2009fk}
  A.~Bazavov {\it et al.}  [MILC Collaboration],
  PoS CD {\bf 09} (2009) 007
  [arXiv:0910.2966 [hep-ph]].



\bibitem{Bazavov:2011fh}
  A.~Bazavov {\it et al.}  [MILC Collaboration],
  PoS LATTICE {\bf 2011} (2011) 107
  [arXiv:1111.4314 [hep-lat]].



\bibitem{Bijnens:2004hk}
  J.~Bijnens, N.~Danielsson and T.~A.~L{\"a}hde,
  Phys.\ Rev.\ D {\bf 70} (2004) 111503
  [hep-lat/0406017].



\bibitem{Bijnens:2005ae}
  J.~Bijnens and T.~A.~L{\"a}hde,
  Phys.\ Rev.\ D {\bf 71} (2005) 094502
  [hep-lat/0501014].



\bibitem{Bijnens:2005pa}
  J.~Bijnens and T.~A.~L{\"a}hde,
  Phys.\ Rev.\ D {\bf 72} (2005) 074502
  [hep-lat/0506004].



\bibitem{Bijnens:2006jv}
  J.~Bijnens, N.~Danielsson and T.~A.~L{\"a}hde,
  Phys.\ Rev.\ D {\bf 73} (2006) 074509
  [hep-lat/0602003].



\bibitem{Dowdall:2013rya}
  R.~J.~Dowdall, C.~T.~H.~Davies, G.~P.~Lepage and C.~McNeile,
  Phys.\ Rev.\ D {\bf 88} (2013) 074504
  [arXiv:1303.1670 [hep-lat]].


\bibitem{Ecker:1989yg}
  G.~Ecker {\it et al.}, 
  Phys.\ Lett.\ B {\bf 223} (1989) 425.



\bibitem{Weinberg:1967kj}
  S.~Weinberg,
  Phys.\ Rev.\ Lett.\  {\bf 18} (1967) 507.



\bibitem{Knecht:1997ts}
  M.~Knecht and E.~de Rafael,
  Phys.\ Lett.\ B {\bf 424} (1998) 335
  [hep-ph/9712457].



\bibitem{Jamin:2001zq}
  M.~Jamin, J.~A.~Oller and A.~Pich,
  Nucl.\ Phys.\ B {\bf 622} (2002) 279
  [hep-ph/0110193].



\bibitem{Pich:2002xy}
  A.~Pich,
  Proc. Institute for Nuclear Theory (World Scientific) {\bf 12}
  (2002) 239 [hep-ph/0205030].



\bibitem{Cirigliano:2003yq}
  V.~Cirigliano, G.~Ecker, H.~Neufeld and A.~Pich,
  JHEP {\bf 0306} (2003) 012
  [hep-ph/0305311].



\bibitem{Bijnens:2003rc}
  J.~Bijnens, E.~G{\'a}miz, E.~Lipartia and J.~Prades,
  JHEP {\bf 0304} (2003) 055
  [hep-ph/0304222].



\bibitem{Das:1967it}
  T.~Das {\it et al.}, 
  Phys.\ Rev.\ Lett.\  {\bf 18} (1967) 759.



\bibitem{Dashen:1969eg}
  R.~F.~Dashen,
  Phys.\ Rev.\  {\bf 183} (1969) 1245.



\bibitem{Donoghue:1993hj}
  J.~F.~Donoghue, B.~R.~Holstein and D.~Wyler,
  Phys.\ Rev.\ D {\bf 47} (1993) 2089.



\bibitem{Bijnens:1993ae}
  J.~Bijnens,
  Phys.\ Lett.\ B {\bf 306} (1993) 343
  [hep-ph/9302217].



\bibitem{Gasser:2003hk}
  J.~Gasser, A.~Rusetsky and I.~Scimemi,
  Eur.\ Phys.\ J.\ C {\bf 32} (2003) 97
  [hep-ph/0305260].



\bibitem{Ananthanarayan:2004qk}
  B.~Ananthanarayan and B.~Moussallam,
  JHEP {\bf 0406} (2004) 047
  [hep-ph/0405206].

\bibitem{Agadjanov:2013lra}
  A.~Agadjanov, D.~Agadjanov, A.~Khelashvili and A.~Rusetsky,
  Eur.\ Phys.\ J.\ A {\bf 49} (2013) 120
  [arXiv:1307.1451 [hep-ph]].


\bibitem{DescotesGenon:2005pw}
  S.~Descotes-Genon and B.~Moussallam,
  Eur.\ Phys.\ J.\ C {\bf 42} (2005) 403
  [hep-ph/0505077].



\bibitem{Sirlin:1981ie}
  A.~Sirlin,
  Nucl.\ Phys.\ B {\bf 196} (1982) 83.



\bibitem{Marciano:1993sh}
  W.~J.~Marciano and A.~Sirlin,
  Phys.\ Rev.\ Lett.\  {\bf 71} (1993) 3629.



\end{thebibliography}

\end{document}